# Nanoscale Charge Transport in Au@PANI Assemblies: Bulk-like Films and Linear Assemblies

*Gyusang Yi* [a,§], *Borja Rodriguez-Barea* [c,d,§], *Gabriele Carelli* [a], *Lukas Mielke* [a], *Andreas Fery* [a,b], *Artur Erbe* [c,d,*], *Hendrik Schlicke* [a,*]

[a] Leibniz Institute of Polymer Research Dresden, 01069 Dresden, Germany.

[b] Chair of Physical Chemistry of Polymeric Materials, Dresden University of Technology, 01062 Dresden, Germany

[c] Institute of Ion Beam Physics and Materials Research, Helmholtz-Zentrum, Dresden-Rossendorf, 01328 Dresden, Germany

[d] Institute of semiconductors and microsystems, Dresden University of Technology, 01062 Dresden, Germany

§ contributed equally to this work

* Corresponding author
E-mail: a.erbe@hzdr.de, schlicke@ipfdd.de

**Abstract**

Hybrid nanostructures from metal nanoparticles equipped with conducting polymer shells are of great interest for use as functional materials in sensing and optoelectronics, as well as for ink-deposited conductors. Here, we investigate the charge transport mechanism of nanostructures composed of gold nanoparticles coated with a polyaniline shell (Au@PANI). In particular, we focus on how geometry influences the charge transport behavior. Highly ordered linear assemblies of Au@PANI nanoparticles were fabricated using template-assisted assembly, while bulk-like films were obtained via drop-casting. Temperature-dependent transport measurements were analyzed using established conductance models. Linear assemblies exhibit more localized transport, characterized by variable-range hopping (VRH) and thermally assisted tunneling (TAT), whereas bulk-like films show more delocalized transport, dominated by Arrhenius-type and thermionic conduction. These findings highlight the critical role of geometry, also due to its effect on electrical field strength in determining charge transport mechanisms in nanoparticle-based hybrid systems.

*Keywords:* Nanoparticles, hybrid materials, nanoelectronics

## 1. Introduction

Designing and controlling charge transport is one of the central challenges in the development of advanced organic and hybrid materials for electronics, sensing, and energy applications. The geometry and dimensional architecture of the active material — whether one-dimensional (1D) nanowires, fibers, or three-dimensional (3D) films and networks — play a critical role in governing electrical properties, mechanisms of conduction, and device performance. Due to advantages in scalability, cost, and power efficiency, deposition and device integration of nanoparticle-based active materials from colloidal solutions via ink-based technologies render a promising approach. Coating and printing techniques have been successfully employed, e.g., for photovoltaic applications[1] and sensing.[2]

In as-synthesized colloidal solutions, nanoparticles are commonly capped with long-chain organic molecules, which act as stabilizers that prevent agglomeration during storage and processing. However, the bulky ligand molecules impede inter-particle coupling in deposited particle layers. Hence, when aiming for electronic applications of materials deposited from such colloidal solutions, modification or post-processing is commonly necessary.[3,4] One widely used approach, especially when aiming for the fabrication of ink-deposited conductive traces from particulate materials, is sintering. Sintering removes the insulating stabilizers between particles and allows them to form a continuous structure, which is mandatory for efficient charge transport. However, it is known to potentially cause the nanoparticles to lose their discrete identities. Thereby, sintering might diminish or even eliminate characteristic features of nanoparticles, such as localized surface plasmon resonance (LSPR) of metal nanoparticles, which is highly dependent on the particle size, shape, and environment.[5] In addition, sintering processes require high temperature, which makes them energy hungry and unsuitable for flexible and heat-sensitive substrates such as polyethylene terephthalate (PET).[6] As an alternative, using core-shell nanoparticles, consisting of an inorganic core surrounded by a conducting polymer shell, as building blocks, is a promising way to overcome those restrictions. The conducting polymer can serve as an electrical bridge between particles, thereby making sintering unnecessary.[7] Additionally, the formed hybrid structures have a high potential for applications in flexible electronics.[8]

Gold nanoparticle (AuNP) cores coated with polyaniline (PANI) shells represent well-known models of such hybrid polymer/metal structures (Au@PANI).[9–14] In this configuration, besides its conductivity, the PANI shell offers sensitivity to environmental pH[10,11,13] and reversible redox activity.[9,10,12] Further, its electrical conductivity can be tuned through protonation or chemical doping.[15] Thus, the combination of pH sensitivity and redox activity of PANI shells with the plasmonic and highly conductive characteristics of AuNP cores offers significant potential as a functional material for (opto)electronic sensing applications.[11,12]

For electronic applications, understanding charge transport mechanisms in materials is a primary step. In a recent study, we investigated the charge transport characteristics of individual Au@PANI particles and how they are affected by chemically modifying their inner interface between the Au core and the PANI shell.[14] In assemblies of nanoparticles, charge transport is strongly dependent on inter-particle interfaces and the assembly geometry. The particulate nature of the building blocks allows shaping the electronic material into arbitrary geometry, even at the nanoscale, by different assembly strategies.

Size and geometry-dependent properties of nanostructures are a key focus in nanoscience, since on the nanoscale dimensionality of materials governs electronic structures and the spatial extension of charge conduction pathways.[16–18] Accordingly, researchers have developed a variety of methods for fabricating one-, two-, and three-dimensional nanoparticle arrays to investigate this dimensional dependence, allowing for systematic study of their physical properties.[19–24] Charge transport in such arrays is strongly dictated by geometry: in 3D bulk-like films, extensive interparticle connectivity enables long-range and thermally activated transport, commonly reporting the highest conductance.[18] 2D planes provide multiple parallel pathways, facilitating percolative transport within the in-plane direction.[25] In 1D wires, electrons are confined to linear chains, where local disorder and electronic localization (e.g., Coulomb blockade) start to dominate.[26] Nonetheless, this strong localization and high aspect ratio allow precise control over transport, making 1D systems particularly suited for nanoscale sensing applications. A mechanistic understanding of charge transport within these varied geometrical frameworks is therefore essential for rationally optimizing the functional performance of materials and for revealing novel electronic phenomena at the frontier of nanoscale science.

Template-assisted self-assembly is a versatile and scalable method for fabricating particle assemblies with highly defined geometries. This method facilitates, e.g., the formation of close-packed particle lines, enabling precise control of line geometry via the shape of the template.[27,28] The geometry of the template is tunable by adjusting fabrication parameters.[29] Previous studies on template-assisted self-assembly have primarily focused on optical investigations of the resulting particle assemblies, such as anisotropic intensity in surface-enhanced Raman spectroscopy (SERS), and chiral properties of more complex, template-generated assemblies.[30–32] However, despite the high potential of core-shell (Au@PANI) nanoparticle-based highly-ordered linear assembly for electronic sensors and optoelectronic applications, their charge transport mechanisms and the effects of geometry remain unexplored.

In this study, we investigate the charge transport mechanisms of two Au@PANI-based assembly types with distinct geometries by comparing nanoscale linear assemblies, having a cross-section of only a few nanoparticles, and their corresponding bulk-like films. The linear assembly structures were fabricated via PDMS template-assisted self-assembly, yielding periodic and line-shaped Au@PANI assemblies with a cross-section of 386 × 344 $nm^2$ deposited on top of electrodes, patterned via electron beam lithography. In contrast, the 3D Au@PANI assemblies were obtained by drop-casting colloidal solutions onto photolithographically patterned electrodes, producing bulk-like films (inhomogeneous and discontinuous). Atomic Force Microscopy (AFM) and scanning electron microscopy (SEM) were employed for the investigation of both structures' morphology. Temperature-dependent transport measurements (50 to 300 K) were conducted and analyzed using different conductance models to disentangle the underlying charge transport mechanisms. This geometry-resolved comparison provides direct insight into how nanoscale confinement and structural organization govern charge transport in hybrid Au@PANI systems.

## 2. Experimental Section

Materials and experimental methods for the fabrication of linear nanoparticle assemblies were based on those described in our previous study.[14]

### 2.1. Materials

Hydrogen tetrachloroaurate trihydrate ($HAuCl_4 \cdot 3H_2O$, 99.99%, abcr GmbH), hexadecyltrimethylammonium chloride (CTAC, 25 wt% in water, Sigma-Aldrich), hexadecyltrimethylammonium bromide (CTAB, 99%, Merck KGaA), L-(+)-ascorbic acid (AA, >99%, Sigma-Aldrich), sodium borohydride ($NaBH_4$, 99%, Sigma-Aldrich), sodium dodecyl sulfate (SDS, >99%, Sigma-Aldrich), aniline (>99%, Sigma-Aldrich), hydrochloric acid (AnalaR Normapur®, 37%, VWR Chemicals), Sylgard 184 PDMS elastomer (Dow Corning). All chemicals were used without further purification. Ultrapure Milli-Q water (18.2 MΩ·cm, pH 8) was used in all aqueous solutions. Thermally oxidized $Si/SiO_2$ (500 nm) wafers were purchased from Silicon Materials e.K., Germany. For metal evaporation, Ti and Au (99.99% pure) were obtained from Kurt. J. Lesker Company GmbH, USA. Positive tone photoresist AZ ECI 3012 and developer AZ 726 MIF were purchased from MicroChemicals GmbH. KI (≥99.0 %), $I_2$ (≥99.8 %), and 2-propanol (≥99.5 %) were purchased from Sigma-Aldrich Chemie GmbH. Ammonium citrate tribasic (>97 %) was purchased from Thermo Fisher Scientific Inc. PMMA A4 resist (950k) was purchased from micro resist technology GmbH, Berlin, Germany. Hydrogen peroxide (30 %) was purchased from Avantor, Inc.

## 2.2. Au nanoparticle Synthesis

Spherical gold nanoparticles were synthesized using a seed-mediated growth approach.[33] Aqueous solutions were used in all steps.

### 2.2.1. Synthesis of 2 nm Au seeds

An aqueous CTAB solution (100 mM, 4.70 mL) with $HAuCl_4$ solution (50 mM, 25 µL) was maintained at 32 °C under vigorous stirring (1400 rpm). $NaBH_4$ (10 mM, 300 µL) was rapidly injected, and the mixture was stirred for 30 s at a vigorous speed. The stirring rate was reduced to 300 rpm, and continued for 30 min at 32 °C. The reaction mixture developed a brownish color, indicating seed formation.

### 2.2.2. Synthesis of 8 nm Au seeds

Aqueous CTAC solution (200 mM, 40 mL) was combined with ascorbic acid solution (100 mM, 30 mL), $HAuCl_4$ solution (0.5 mM, 40 mL), and 1 mL of the 2 nm seed solution. The mixture was stirred at 300 rpm for 15 min. The resulting particles were purified by two centrifugation and redispersion cycles (15000 rcf, 1 h) and resuspended in CTAC solution (20 mM, 10 mL).

### 2.2.3. Synthesis of 50 nm Au@CTAC nanoparticles

One solution containing CTAC (60 mM) and ascorbic acid (1.3 mM) in water (200 mL) was prepared. An aliquot of the 8 nm seed solution (1.71 mL) was added to this solution under stirring at 380 rpm. Separately, a growth solution ($HAuCl_4$: 1 mM; CTAC: 60 mM; 200 mL) was preheated in a 45 °C water vessel. Three 60 mL solutions prepared by syringes were introduced dropwise into the seed solution using a syringe pump (0.5 mL/min). After syringe addition, the residual growth solution ($HAuCl_4$: 1 mM; CTAC:

60 mM; 20 mL) was added to the seed solution to promote etching, and the mixture was stirred for a further 15 min. The final product was purified by two centrifugation and redispersion cycles in CTAC (10 mM) at 2000 rcf for 20 min. The Au@CTAC solution was kept in CTAC (10 mM, 30 mL).

### 2.3. PANI Shell Coating (Au@PANI)

Polyaniline shells were coated on the Au nanoparticles by the surfactant-assisted oxidative polymerization.[34] Au@CTAC nanoparticle dispersions (0.5 mg/mL, 9.87 mL), Milli-Q water (44.46 mL), SDS (80 mM, 4.33 mL), HCl (1 M, 296 µL), and aniline (10 mM, 5.93 mL) were added sequentially. Lastly, APS (13.33 mM, 7.405 mL) was injected under vigorous stirring (900 rpm). The mixture was stirred at this speed for 1 h, after which the stirring speed was reduced to 130 rpm. The reaction was continued until the solution turned green—approximately 18 h.

### 2.4. Template fabrication

Template fabrication followed a previously reported procedure.[29] PDMS was prepared from the Sylgard kit by mixing cross-linker (4.58 g) and prepolymer (22.92 g) at a 1:5 (w/w) ratio. It was transferred into a square polystyrene petri dish at the even level and allowed to rest at room temperature for 24 h, then cured at 80 °C for 5 h, followed by overnight storage at room temperature. The cured PDMS was cut into strips measuring 1 × 4.5 $cm^2$ and elongated by 60% using a stretching device. The stretched strips were subjected to oxygen plasma treatment (80 W, 5 min, 0.1 mbar) and released from the applied strain, producing wrinkled surfaces. The wrinkled PDMS template was cut into 1 × 1 $cm^2$ pieces and rinsed with distilled water and dried with $N_2$ gas prior to use.

### 2.5. Electrode Fabrication

#### 2.5.1. Electron beam lithographic Fabrication of Interdigitated Electrodes Microstructures

Device layout of electrodes with a 350 nm inter-electrode distance was designed and patterned on $SiO_2$ substrates via electron beam lithography (Raith e-line Plus). Specifically, p-Si (100) wafers with a 280 nm thermally grown oxide layer served as the insulating substrate. Prior to processing, the substrates were ultrasonicated in acetone for 10 min and in 2-propanol (IPA) for 2 min. A single layer of PMMA-A4 resist was spin-coated for 60 s and baked at 150 °C for 10 min. Electron beam exposure was carried out at an acceleration voltage of 10 kV, using a 30 µm aperture and a dose of 100 µC/cm². The resist was developed sequentially in IPA/DI water (7:3) and DI water, each for 30 s. Creavac CREAMET 600 was used to deposit a 5 nm layer of Ti evaporated at a rate of 2 Å/s, followed by a 50 nm Au layer evaporated at a rate of 5 Å/s. The process was completed by an overnight lift-off in acetone.

#### 2.5.2. Photolithographic Fabrication of Interdigitated Electrodes Microstructures.

Si/$SiO_2$ wafers were treated in an oxygen plasma for 3 min. A layer of Ti/Au (10 nm/40 nm) was deposited using a Telemark e-beam source. Chips featuring twelve interdigitated electrode (IDE) structures (15 µm channel length, total overlap width of the IDE fingers of 17.16 mm) were then fabricated photolithographically. The substrates were first treated in an oxygen plasma for 3 min and then baked at 200 °C for 10 min to remove adsorbed water. Afterwards, the substrates were coated with positive-tone photoresist AZ ECI 3012 on a spin-coater (60 s, 600 rpm/s, 6000 rpm) and soft-baked at 90 °C for 60 s. This was followed by an exposure to UV light for 3.3 s (~16 mW/cm$^2$) through a custom-designed photomask using a Karl Suss MJB-3 UV 300 mask aligner. The post-exposure bake was performed at 110 °C for 60 s. The resist was developed for 60 s by immersion in an AZ 726 MIF developer, followed by rinsing the substrates with water and drying them in a nitrogen stream.

Gold etching was performed by immersing the substrates in a gold etchant solution ($KI/I_2/H_2O$, 4/1/400, $m/m/m$) for 2.5 min and subsequent rinsing with water. The residual photoresist was removed by exposure to UV light for 10 s with the aforementioned mask aligner and subsequent development with an AZ 726 MIF developer for 2 min. Titanium etching was performed by immersion in a 5 $wt\%$ triammonium citrate solution in hydrogen peroxide (30 %) for 135 s at 35 °C, followed by thoroughly rinsing the substrates with water and drying them in a nitrogen stream. The substrates were additionally cleaned by subsequent immersion in acetone and 2-propanol. Prior to deposition of the material to be tested, the electrode structures were again treated with an oxygen plasma for 3 min.

## 2.6. Assembly of Au@PANI Nanoparticles

### 2.6.1. Fabrication of Colloidal Nanowires

Template-assisted assembly of nanoparticles was carried out via spin-coating on wrinkled PDMS substrates.[27] The wrinkled templates were hydrophilized via oxygen plasma treatment (80 W, 45 s, 0.2 mbar). After casting Au@PANI solution (5 µL, 20 mg/mL, SDS 2 mM) onto the hydrophilized templates, the spin-coating was conducted in two steps: 222 rpm for 2 s and 1700 rpm for 90 s (Headway Research Inc.). Electrodes patterned by electron beam lithography were cleaned via sequential sonication (80 kHz) in acetone and isopropanol (5 minutes each) and subsequently hydrophilized by oxygen plasma treatment (80 W, 3 min, 0.2 mbar). Milli-Q water (10 µL) was spread across the hydrophilized wafers, and nanoparticle-coated PDMS stamps were placed in contact with a 1 kg pressure overnight to transfer the nanoparticle arrays onto electron beam lithographically patterned electrodes.

#### 2.6.2. Bulk-like Film Deposition of Au@PANI Nanoparticles

Photolithographically patterned interdigitated electrodes were immersed sequentially in acetone and isopropanol for 30 min each. After immersion, the surface was hydrophilized by oxygen plasma treatment (80 W, 3 min, 0.2 mbar). Au@PANI solution (2 µL, 4 mg/mL, SDS 2 mM) was drop-cast on the photolithographically patterned electrode and dried overnight.

### 2.7. Extinction Spectroscopy (Vis/NIR Region)

Solution-phase extinction spectra of nanoparticle solutions were recorded using a Cary 5000 spectrometer (Agilent, USA) in a wavelength range of 400–1000 nm at a scan rate of 600 nm/min, with a fixed spot size of $3 \times 4\ mm^2$. 2 mL of Milli-Q water was measured as baseline. 2 µL of colloidal solution was diluted in 2 mL of Milli-Q water for measurement.

### 2.8. Transmission Electron Microscope (TEM)

TEM imaging was performed on a Libra 120 microscope (Zeiss, Germany) operating at 120 kV. Samples (around 0.1 mg/ml, 5.5 µL in Milli-Q water) were drop-cast onto CF200-Cu-50 TEM grids (carbon support film on 200 mesh copper, Electron Microscopy Sciences, USA) and dried overnight. Image analysis was conducted using ImageJ software.

## 2.9. Atomic Force Microscopy (AFM)

The morphology of colloidal nanowires and PDMS templates was characterized using an atomic force microscope (AFM; Dimension ICON, Bruker, Billerica, MA, USA). ScanAsyst-Fluid+ probes (tip radius: 2 nm; spring constant: 0.7 N/m; resonance frequency: 150 kHz; silicon nitride cantilevers) were used to scan the surfaces of the templates and colloidal nanowires. AFM data were analyzed using Gwyddion 2.58 software.[35]

## 2.10. Scanning electron microscope (SEM)

Scanning electron microscopy (SEM) for morphological characterization was carried out using the EBL (Raith e-line Plus) operated in SEM mode. Substrates were mounted with carbon tape to reduce charging effects, and imaging was performed at an accelerating voltage of 10 kV using the InLens secondary electron detector, a 30 µm aperture and a working distance of 9.5 mm.

## 2.11. Optical Microscopy

High-resolution microscope images were obtained using an optical microscope (Nikon ECLIPSE LV100ND). Images were captured with NIS-Elements AR software (version 4.20.01).

## 2.12. Temperature-dependent electrical characterization

Electrical characterization was performed via current–voltage (I–V) measurements: bulk-like Au@PANI films were measured in a two-terminal configuration, while linear assemblies were characterized using both two- and four-terminal measurements under darkness and high vacuum (base pressure ~ $10^{-5}$ mbar) using a Keithley 4200A-SCS parameter analyzer. Two (or four) tungsten probe tips of 25 μm were positioned on the gold contacts. Temperature-dependent measurements were carried out in a liquid-helium continuous-flow cryostat. Samples were cooled to 50 K, and data were recorded at temperatures increasing in steps of 25 K with a base pressure of ≈ $10^{-7}$ mbar. I–V sweeps for the two-terminal devices were performed from 0 to +3 V, then to −3 V, and back to 0 V. Instead, for the four-terminal devices, current sweeps from 0 to +100 nA, then to -100 nA, and back to 0 nA. Resistances of the bulk-like films were obtained from linear fits to the I–V curves. For the linear assembly, which exhibited non-linear behavior, the resistance was evaluated at different input currents (0, 15, 50, and 100 nA), where zero-bias conductance was evaluated via numerical differentiation of the IV data around $V_{sd}$=0 (source-drain voltage).

## 3. Result and discussion

### 3.1. Fabrication of Au@PANI assemblies

Initially, a seed-mediated growth method[33] was adapted to synthesize spherical, hexadecyltrimethylammonium chloride (CTAC) capped AuNPs by the reduction of tetrachloroauric acid (Fig. 1a). CTAC stabilizes gold nanoparticles via formation of bilayers in aqueous solution.[14] In this structure, chloride ions interact with the gold core, as confirmed by SERS in previous research,[14] while the positively charged hexadecyltrimethylammonium groups form the outer layer, other chloride ions serve as counterions near the outer positively charged layer (hexadecyltrimethylammonium), generating electrostatic repulsion between particles and thereby preventing aggregation. A representative TEM image and core diameter distribution of Au@CTAC are shown in Fig. 1b and 1d, respectively. The average particle diameter is 54.5 ± 3.8 nm.

To grow a PANI shell, a surfactant-assisted oxidative polymerization method [14,34] was employed (Fig. 1a). Aniline monomers were encapsulated with sodium dodecyl sulfate (SDS) micelles. These negatively charged micelles replaced the negatively charged chloride ions of Au@CTAC, enabling the introduction of aniline to the AuNP surface. Anilines were polymerized with the oxidant, ammonium persulfate (APS), under acidic conditions (pH ~ 2.3), to form a shell surrounding the AuNP surface.[14,34] TEM analysis (Fig. 1c) confirmed shell formation, with an average shell thickness of 20.7 ± 4.2 nm (Fig. 1d).

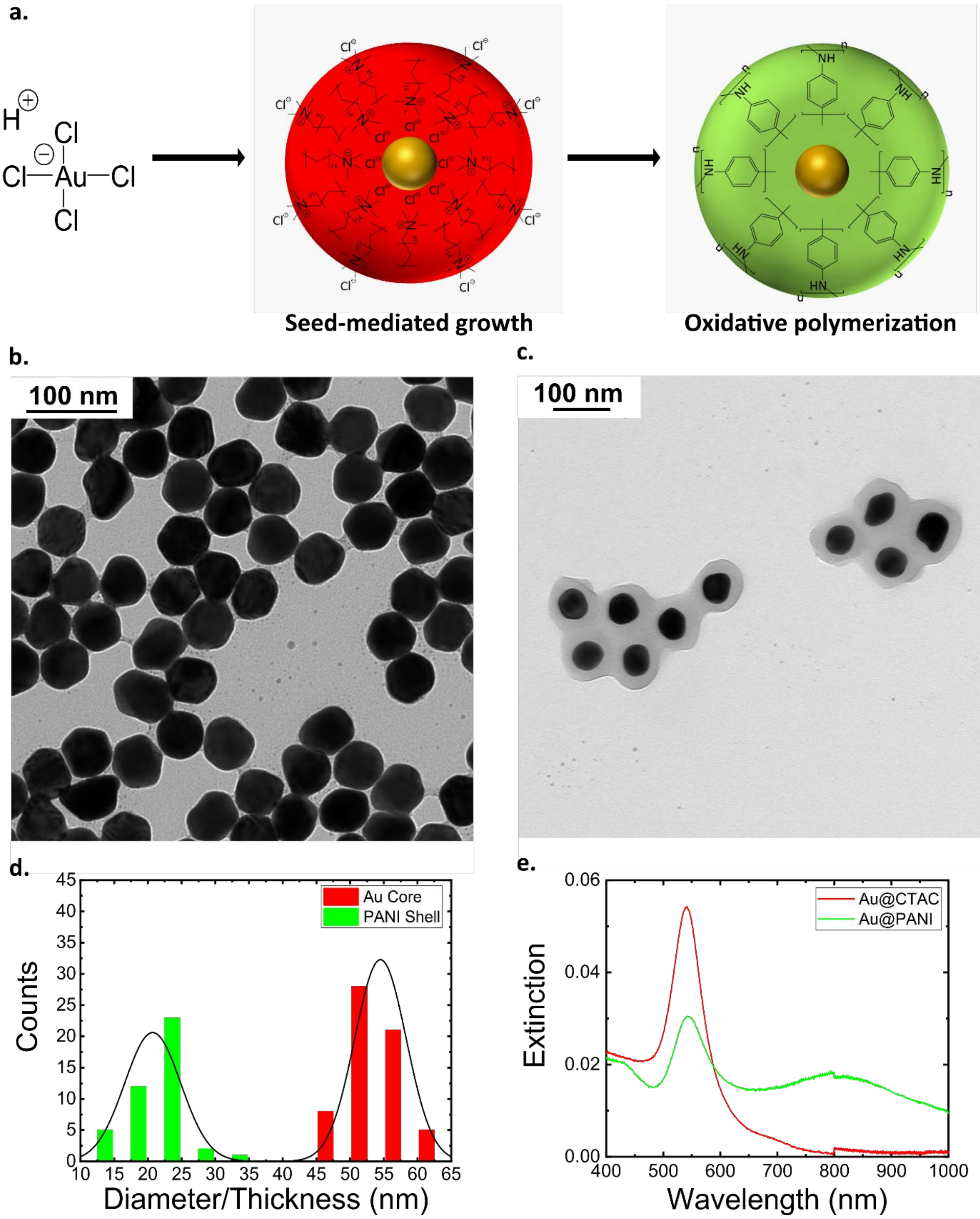

**Fig. 1.** Particle Characterization: a) Schematic illustration of CTAC-stabilized gold nanoparticles (Au@CTAC) obtained via a seed-mediated growth process and gold/PANI core/shell nanoparticles (Au@PANI) synthesized via the surfactant-assisted oxidative polymerization, b,c) TEM image of b) Au@CTAC, and c) Au@PANI nanoparticles. d) Size distribution of Au core diameter and PANI shell thickness of representative core (Au core, red), and shell (PANI, green) samples. e) Extinction spectra Au@CTAC and Au@PANI nanoparticles.

After shell formation, a slight red shift of the LSPR maximum from 542 nm (Au@CTAC) to 543 nm (Au@PANI) became visible (Fig. 1e). This shift is attributed to the PANI shell altering the local refractive index and increasing the overall particle size, leading to the change of dielectric response of particles.[5,14,36,37] The PANI shell also exhibits a broad absorbance band centered around 820 nm, whose position depends on the protonation state of PANI.[10,14]

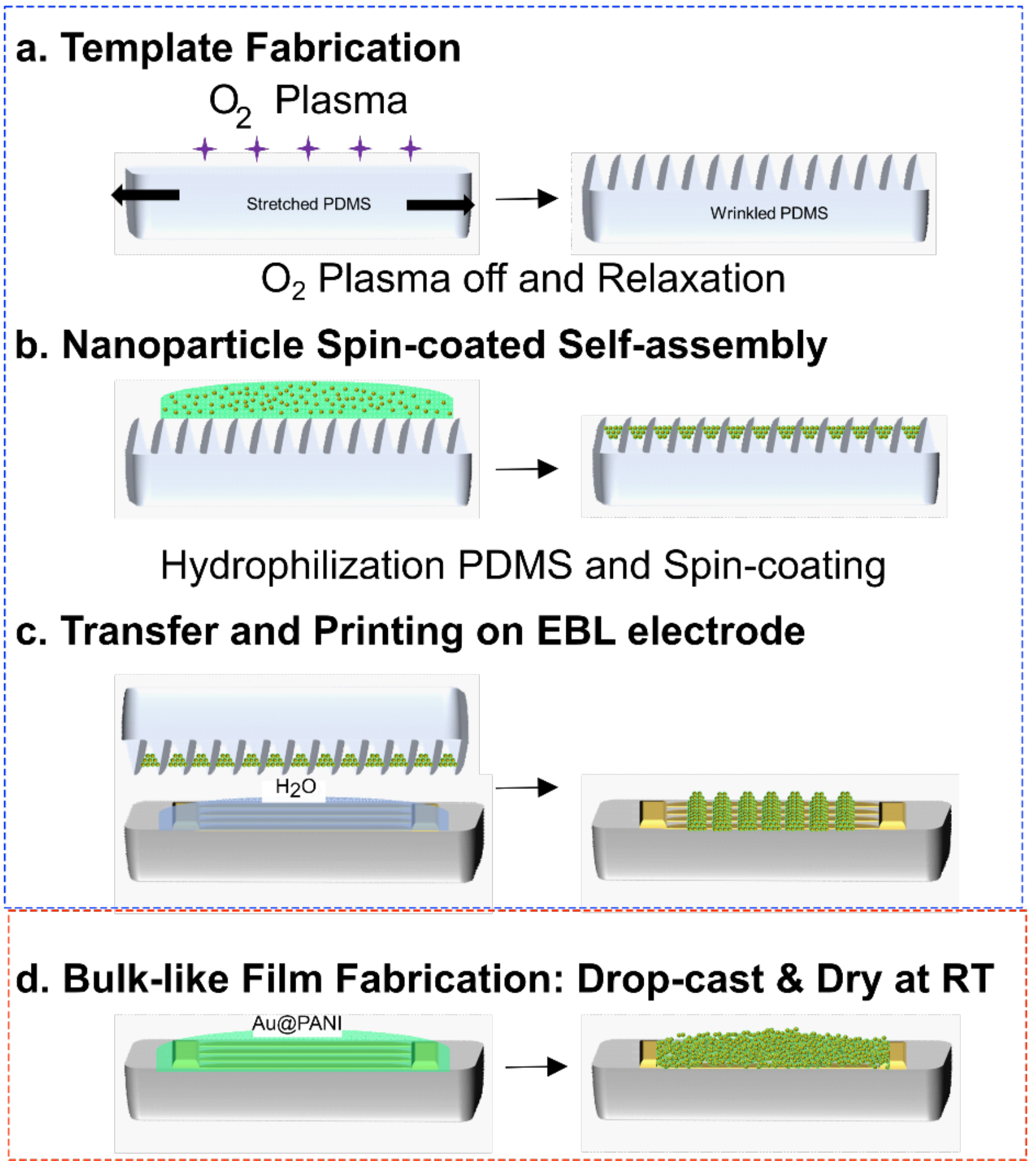

**Fig. 2.** Fabrication of linear assembly via template-assisted assembly (blue box): a) Template fabrication, b) Spin-coating of colloidal Au@PANI nanoparticles onto the template, and c) transfer of the nanoparticle linear assembly onto electrodes. d) Bulk-like film fabrication via drop-casting at room temperature (red box).

For the formation of linear assemblies, a template-assisted approach was chosen (Fig. 2, blue box).[27] Initially, oxygen plasma treatment was employed to modify the surface properties of a strained PDMS substrate. Subsequently, release of strain induced the formation of nano- to microscale wrinkles at the surface of these bilayer structures (cf. Fig. 2a).[38] Wrinkled templates were fabricated using tunable parameters, such as varying oxygen plasma treatment intensity and extent of strain application to adjust wavelength and amplitude.[29] Second, the PDMS template was hydrophilized via oxygen plasma treatment, and nanoparticles were applied via self-assembly in the template's valleys after deposition of their colloidal solution via spin-coating. Following self-assembly of the particles in the PDMS wrinkles, they were transferred onto electrode microstructures, which were pre-patterned via electron beam lithography. These structures have an electrode-electrode distance of 350 nm and consist of a 5 nm Ti adhesion layer, deposited via e-beam evaporation, and a 50 nm Au layer, deposited by thermal evaporation. A droplet (10 µL) of water (Milli-Q) was dispensed onto the hydrophilized electrode surface after oxygen plasma treatment, and the template carrying the aligned linear Au@PANI assemblies was placed onto the electrode substrates for transfer. Applying a pressure of ~ 1 bar by placing a weight on the top of the PDMS template enhanced the transfer efficiency, resulting in precise and reliable transfer of the linear assemblies onto the electrodes after removal of the PDMS templates. Fig. 3 depicts topographic AFM scans and extracted height profiles of the obtained colloidal nanowires and the PDMS template used for fabrication.

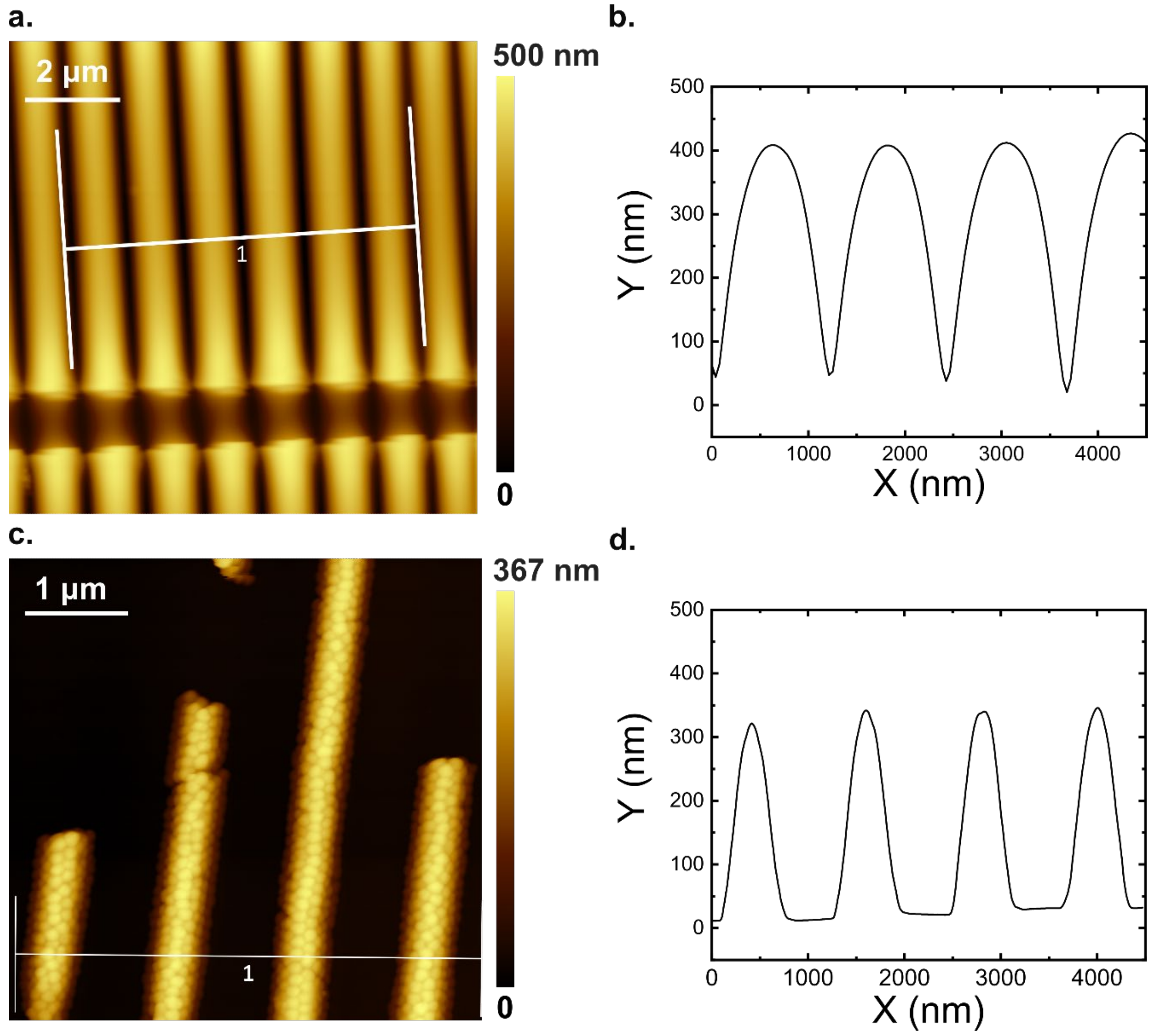

**Fig. 3.** a) Topographic AFM scan of a representative PDMS template's wrinkled surface. b) Line profile extracted from a. c) Topographic scan of Au@PANI linear assemblies fabricated using the template shown in a. d) Height profile of the representative Au@PANI linear assemblies depicted in c.

For a representative sample shown in Fig. 3a, AFM analysis revealed the template's wrinkle wavelength of 1236 ± 53 nm. This value matches the periodicity of the corresponding linear Au@PANI assemblies (1198 ± 23 nm) deposited using this template within a 3% margin of error. Fitting of Gaussian Functions to the wrinkle profiles was used to extract the full width at half maximum (FWHM) and height values (Fig. S2-S3). The FWHM of the template valleys is 386 ± 6 nm, while that of the particle lines is 372 ± 7 nm, also within a 3% deviation. The template depth (351 ± 12 nm) corresponds to the linear assembly height (344 ± 9 nm) within a 2% margin. These consistent measurements show that our approach produces highly organized, close-packed particle-based linear assemblies.[27] Furthermore, the linear assembly geometry is easily tunable by adjusting template fabrication parameters, like plasma treatment conditions and applied strain.[29]

3D (bulk film) structures were fabricated by direct drop-casting of Au@PANI solutions onto photolithographically patterned electrodes, as schematically shown in Fig. 2d (red box). The photographically patterned interdigitated electrodes consist of Ti/Au with 15 µm channel length and total overlap width of the interdigitated electrode fingers of 17.16 mm. Before drop-casting, the surface of the electrodes was hydrophilized via oxygen plasma treatment, which helped the droplet of colloidal nanoparticle solution spread well on the surface. The droplet was dried overnight at room temperature. Corresponding SEM images of linear assemblies and bulk-like films are shown in Fig. 4.

### 3.2. Characterization of Charge Transport Properties

To investigate the geometry-dependent electronic properties of the produced Au@PANI assemblies, current-voltage (*I*-*V*) measurements were performed in darkness under high-vacuum (base pressure: $1x10^{-5}$ mbar). Under these conditions, both line and bulk systems exhibited stable and reproducible characteristics. For the linear assemblies, four-terminal configurations were employed to separate the contact resistance contribution from the intrinsic material resistance. Two-terminal measurements on interdigitated electrodes were conducted for bulk-like film assemblies. In these film-like structures, the current distributes across many junctions and a broad electrode area, effectively parallelizing the pathways and enabling more reproducible two-terminal measurements compared to the lines. Note that all electrical characterization was performed prior to SEM imaging, since exposure to the electron beam is known to damage PANI fibers and significantly reduce their electrical conductivity.[39] This effect was also observed in our study: a representative example of the characteristics before and after e-beam exposure is provided in Fig. S9.

#### 3.2.1. Three-dimensional Au@PANI Films

The scanning electron microscope (SEM) image presented in Fig. 4a and a corresponding image measured on a second electrode structure in Fig. S4a depict a three-dimensional gold–polyaniline (Au@PANI) composite film covering the two interdigitated gold electrodes, separated by 15 µm. The image indicates that microscopically, the film is almost defect-free. Unlike in other systems,[40,41] we observe a homogeneous and dense integration of uniform metallic nanoparticles within the polymer matrix, clearly visible in the SEM image due to the contrast differences between the two materials in the composite. This underlines the advantage of employing highly defined core-shell nanoparticles as building blocks for the fabrication of organized 3D nanocomposites. The integration of metallic nanoparticles within the conducting polymer is expected to act as charge acceptors and localized conduction centers, facilitating long-range charge transport between the positively charged nitrogen sites in polyaniline and negatively charged gold nanoparticles.[42] On large scales, however, inhomogeneities cannot be avoided, as shown in Fig. S1. This poor macroscopic distribution is mainly caused by the lack of control in the drop casting process.

As shown in Fig. 4b, the *I*–*V* curves measured on bulk-like films exhibit a linear behavior across the entire measured bias range of ±3 V, corresponding to an applied field strength of $2x10^5$ V/m for the largest voltage. The linearity is preserved over the full temperature range investigated (50–300 K), indicating ohmic transport dominated by bulk percolation pathways within the PANI–AuNP network. The observed temperature dependence of the conductance suggests thermally activated charge transport, consistent with conduction in disordered polymer–nanoparticle composites, while no field-induced non-linearity is detected within the accessible electric-field regime.

### 3.2.2. Linear Au@PANI assemblies

The scanning electron microscope image in Fig. 4c depicts a single linear Au@PANI assembly (resembling a quasi-1D nanowire) bridging four bottom electrodes, which are positioned equidistantly with a separation of 350 nm. This assembly exhibits a linear geometry with negligible branching, reflecting the uniformity achieved by the template-assisted process. Its surface morphology reveals densely packed gold nanoparticles uniformly embedded along with the polyaniline, forming a continuous hybrid structure. The clear contrast in the SEM image highlights both the smooth polymer coating and the uniform incorporation of the Au nanoparticles, underlining a well-defined architecture.

Electrical transport properties of two individual nanowire-like systems are shown for the four-terminal configuration, one in the main text (Fig. 4d) and a repeated measurement on a second electrode in the SI (Fig. S4d), to highlight reproducibility of the results. Prior to the four-terminal measurements, two-terminal measurements were carried out to confirm a stable connection between all four electrodes to ensure a reliable current injection when measuring the voltage drop across the two internal electrodes (Fig. S10). As depicted in Fig. 4d, the obtained *I-V* curves show a non-linear behavior, indicating that the charge transport deviates from purely metallic conduction or linear tunneling due to interactions within the confined linear chains. In this context, transport was investigated under four current-bias regimes: zero bias (0 nA), low bias (15 nA), medium bias (50 nA), and high bias (100 nA). The zero-bias regime is particularly critical to distinguish between influences of the electric-field-on charge transport mechanisms and effects arising purely from device geometry. The zero-bias conductance was extracted from the differential resistance $dI/dV$ evaluated in the vicinity of source-drain voltage $(V_{\mathrm{sd}}) = 0$ V. Analysis of differential conductance, shown in Fig. 5, reveals the absence of a conductance suppression or transport gap around zero bias, indicating that charge injection is not inhibited within the applied field range (>1x$10^6$ V/m). Notably, despite the finite zero-bias conductance, the *I*–*V* characteristics remain non-linear over the entire explored voltage range. This behavior contrasts sharply with that observed in the 3D film assemblies, which exhibit linear, ohmic transport under comparable bias conditions. The pronounced non-linearity in the linear devices is therefore attributed to electric-field-assisted transport mechanisms that become operative due to the substantially enhanced local electric fields at short electrode separations, rather than to geometric confinement alone.

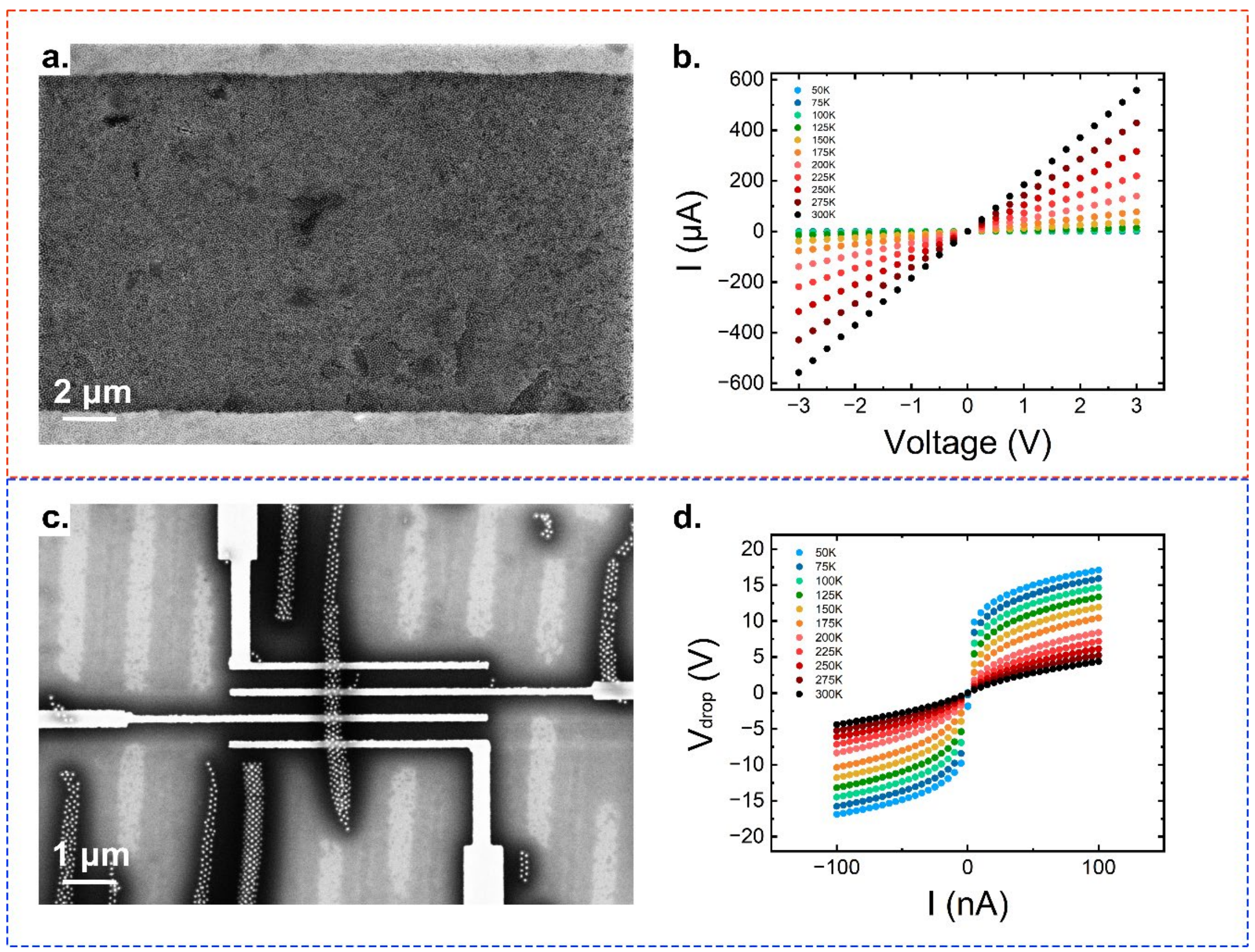

**Fig. 4.** a) An SEM image of the Au@PANI 3D bulk-like film in a section of the IDE electrodes b) the two-probe electrical characterization across the 50–300 K temperature range (red box), c) an SEM image of the PDMS-template-assisted Au@PANI linear assemblies positioned on top of four bottom electrodes, and d) the four-probe electrical characterization in the 50–300 K temperature range (blue box)

Notably, non-linear *I–V* behavior has previously been observed for field strengths of $10^7$ V/m in a vertical device architecture, for AuNP grown in situ within PANI nanofibers via a redox reaction with chloroauric acid.[43] Our study reveals a clear distinction between line and film composite geometry, despite using the same nanoparticle building block. In quasi-1D linear nanostructures, consisting of discrete and linearly arranged Au@PANI nanoparticles, the number of available pathways between the nanoparticles is limited, forcing anisotropic charge transport along the length of the confined structure. Under these higher-field conditions, charge transport becomes dominated by field-assisted processes at interparticle junctions, such as barrier lowering and hopping enhancement, which are intrinsically non-linear. The quasi-1D confinement amplifies the impact of local heterogeneities, including variations in interparticle spacing or polymer coating thickness. By contrast, the 3D networks provide numerous interconnected pathways, allowing current to preferentially flow through the most conductive routes, yielding linear *I-V* characteristics. This demonstrates that, even with identical nanoparticle building blocks, geometry can dictate the dominant charge transport mechanism and tune the conduction in hybrid materials.

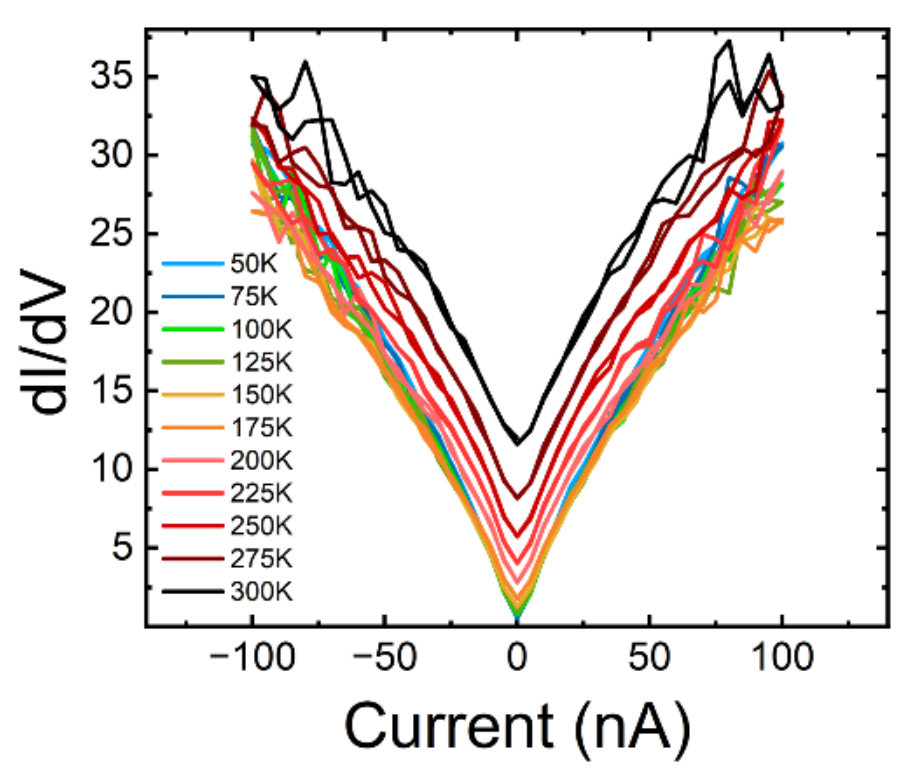

**Fig. 5.** Differential conductance of the linear chain assembly shown in Fig. 4c from a four-terminal configuration at different temperatures ranging from 50K–300 K.

### 3.2.3. Au@PANI conductivity

The reported electrical conductivity of polyaniline (PANI) nanostructures varies significantly, as evidenced by direct literature comparisons between 3D films and 1D fibers. For example, X. Zhang et al. (2004)[44] reported that the conductivity of polyaniline, in its pressed-pellet form, is typically in the range 2–10 S/cm. The conductivity of nanocomposites incorporating gold nanoparticles (26 nm size) was 0.3 S/cm,[45] and expected to depend on the nanoparticle concentration. E. K. Mushibe, et.al. (2012) studied bare PANI fibers with an electrical conductivity of 4.2 S/cm, which, after the introduction of AuNPs, increased to 34.0 S/cm.[46] Due to the inhomogeneity in the 3D Au@PANI bulk film thickness, we report and discuss the electrical conductance rather than the conductivity of our systems. We present the conductance values for both of our systems in Fig. S8, displaying an expectedly significantly lower value for the reduced geometry architecture.

### 3.2.4. Charge Transport Mechanisms in Au@PANI

The charge transport mechanism in both the bulk and linear Au@PANI assemblies was investigated over the 50–300 K temperature range and is presented in Fig. 6. First, an Arrhenius analysis displayed in Fig. 6a reveals distinct behaviors for the two systems. In both cases, the change in slope suggests that more than a single dominant mechanism governs charge transport across the different temperature regimes. The linear system exhibits a more pronounced slope that gradually approaches a plateau across all four bias conditions. Notably, the bias dependence indicates that lower fields produce behavior more akin to bulk-like systems, although this effect is not expected to be purely field-induced. On the other hand, the bulk-like system displays an almost linear Arrhenius trend, consistent with a more conventional thermally activated transport mechanism.

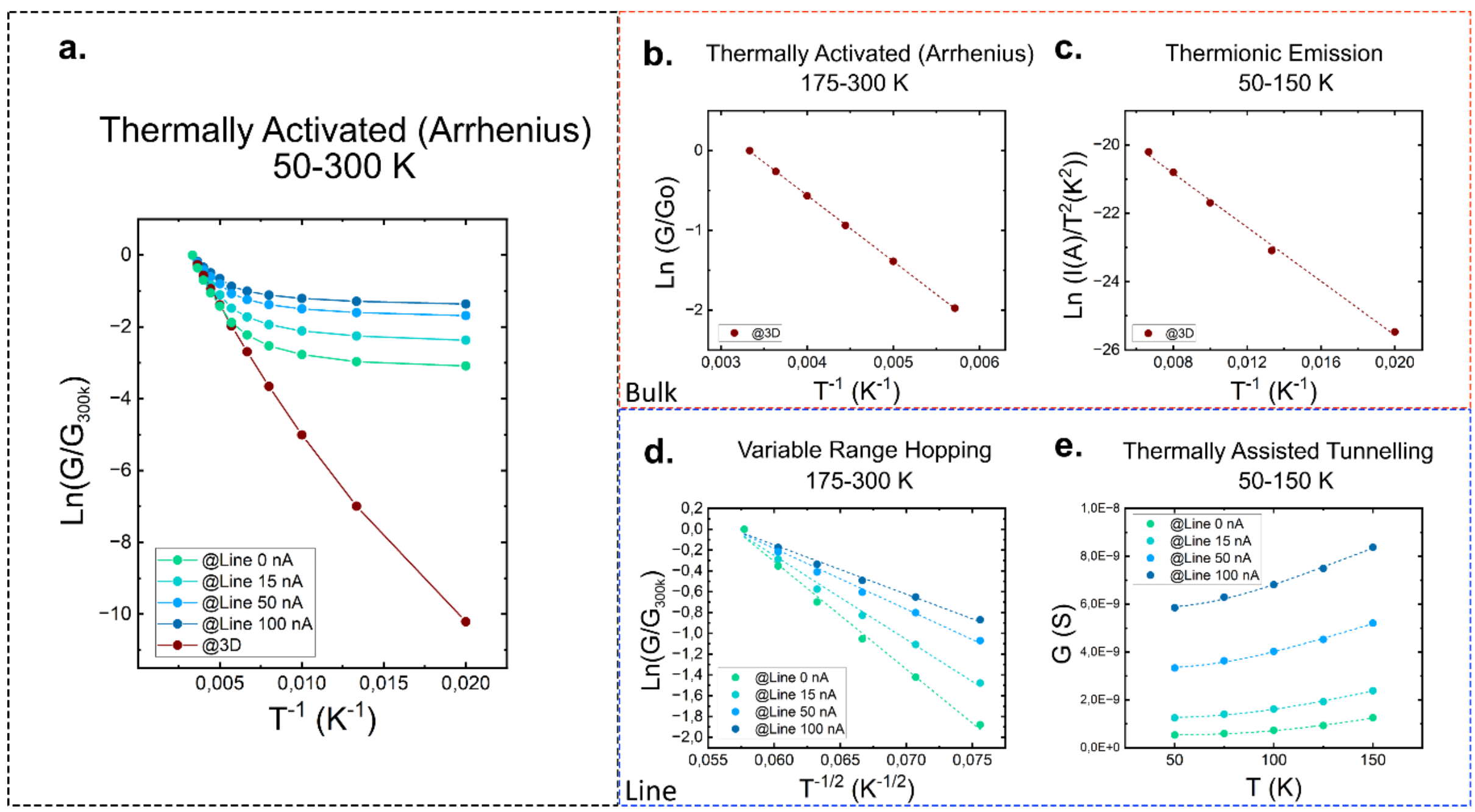

**Fig. 6.** a) Logarithm of the normalized conductance plotted against $T^{-1}$ for linearization according to the Arrhenius mechanism across the 50–300 K temperature range for both linear and bulk-like systems. The solid line serves as a visual guide. The top row (red box) shows the fittings for the bulk-like film: b) Arrhenius plot restricted to the high-temperature regime (175–300 K), with a linear fit shown as a solid dashed line. c) Displays thermionic conduction mechanism in the low-temperature regime (50–150 K), with a linear fit shown as a solid dashed line. The bottom row (blue box) shows the fittings for the line assembly: d) 1D VRH model in the high-temperature regime (175–300 K), with a linear fit shown as a solid dashed line. e) Thermally Assisted Tunneling (TAT) model in the low-temperature regime (50–150 K), with a linear fit shown as a solid dashed line.

Now, the charge transport in Au@PANI films was analyzed over two temperature regimes. In the 175–300 K range (Fig. 6b), an Arrhenius-type analysis, $\ln\frac{G}{G_{300\,\mathrm{K}}} = -\frac{E_\mathrm{a}}{k_\mathrm{B}T}$ yields an activation energy $E_\mathrm{a}$ = 71 meV. This value is similar to the measured activation energy (74.4 meV) of HCl-doped PANI in literature; [47] thus, indicating that PANI-PANI coupling is the main contributor of charge transport.

At lower temperatures (50-150 K) in Fig. 6c, thermionic emission provides the best fit following the equation: $\ln\left(\frac{I}{T^2}\right) = \ln A_\mathrm{G} - \frac{\Phi}{k_\mathrm{B}T}$ which is consistent with a more localized conduction process dominated by well-defined interparticle junctions, specifically those at PANI-PANI interparticle junctions. This transition in transport mechanism arises when thermal energy falls below a critical threshold, highlighting the role of PANI–PANI interparticle interfaces in mediating low-temperature charge transport.

The charge transport in the linear Au@PANI-based system was investigated across the same temperature ranges. In the high-temperature regime of 175–300 K (Fig. 6d), conduction is best described by variable-range hopping (VRH). The VRH for one-dimensional transport is appropriate for the present nanoscale system, given the lateral dimensions of ~380 nm. From the linear fit in the $T^{-1/2}$ representation, the lowest characteristic temperature $T_0$ was observed for the highest current input, yielding a value of 2.2x10$^3$ K. This is significantly lower than the average $T_0$ of 5.5x10$^3$ K reported for polyaniline-Au nanorods composited by Gupta et al.[48] The lower $T_0$ is associated with more compact structures, indicating that the present hybrid system promotes denser packing and leads to more effective PANI-PANI coupling. Furthermore, as the input current decreases, $T_o$ increases, obtaining a value of 3.4x10$^3$ K for 50 nA, 6.5x10$^3$ K for 15 nA, and 10x10$^3$ K for 0 nA. This trend suggests that at lower biases, corresponding to a field strength of ~3 × 10$^6$ V/m, the coupling between PANI shells is weaker, whereas at higher biases, with a field strength of ~10$^7$ V/m, the voltage supports transport between the shells by overcoming transport barriers.

At lower temperatures in the 50-150 K range (Fig. 6e), VRH becomes inefficient and transport is dominated by thermally assisted tunneling (TAT) through short PANI-PANI junctions. Non-linear fits are performed to extract the relevant energies using: $G = G_{base} + G_1 \cdot e^{(-\frac{E_a}{K_B T})}$, where $G_{\text{base}}$ is the temperature-independent baseline conductance, $G_1$ is a prefactor for the thermally activated process and $E_a$ is the activation energy. The fitted values, included in Table S1, show negligible dependence on the applied current, as tunneling is governed primarily by barrier properties rather than thermal or electrical excitation. However, with higher bias, the temperature-independent conductance increases, consistent with enhanced carrier injection.

Overall, our results demonstrate that charge transport in the linear assemblies is dominated by VRH at high temperatures (175–300 K), with characteristic $T_0$ values lower than those reported for similar systems in literature,[48,49] whereas the bulk-like system exhibits more conventional thermally activated conduction. This highlights the critical role of nanoscale geometry in determining charge transport trajectories, while emphasizing the influence of local field strength. Both together shape the electronic behavior of the hybrid assemblies.

## 4. Conclusion

In this study, we investigated the charge transport behavior of Au@PANI hybrid nanostructures, focusing on the influence of the geometry of the assembled structures. Bulk-like assemblies obtained by drop-casting show enhanced delocalization, with linear *I*–*V* behavior and conduction dominated by Arrhenius-type and thermionic mechanisms. Highly ordered linear assemblies, prepared via template-assisted assembly, exhibit non-linear *I*–*V* characteristics and conduction governed by variable-range hopping (VRH) and temperature assisted tunneling (TAT). Both systems display a transition of charge transport mechanism in the 150–175 K range, corresponding to a crossover between localized and thermally activated transport regimes. The distribution of the electric field strongly depends on the microscopic geometry; therefore, the effects of higher fields can be seen in the linear assemblies. These results demonstrate that the geometry of nanoparticle assemblies governs electron localization and the dominant charge transport mechanisms in nanoscale systems. Importantly, these Au@PANI hybrids enable efficient electron transport without high-temperature sintering, while preserving structural integrity and allowing integration into flexible devices whose geometry can be tailored for specific electronic functions.

**Acknowledgment**

G.Y., B.R.B., L.M., A.F., A.E. and H.S. acknowledge financial support funded by the Deutsche Forschungsgemeinschaft (DFG, German Research Foundation) – GRK 2767 – Project number 451785257. The support by the Nanofabrication Facilities Rossendorf (NanoFaRo) at the IBC is gratefully acknowledged. The authors thank Fiona Tenhagen for supporting in photolithographic fabrication of electrode microstructures.

**Author contributions**

Conceptualization: A.F., A.E. and H.S.; Particle Synthesis and characterization: G.Y.; Assemblies Fabrication: G.Y; Electrical Measurement and Charge Transport Mechanism Modeling: B.R.B.; Electrode Fabrication: B.R.B., G.C, and L.M.; Writing manuscript draft: G.Y., B.R.B.; Writing review and editing; G.Y., B.R.B., G.C., L.M., A.F., A.E. and H.S. Supervision: A.F., A.E. and H.S.; Funding acquisition: A.F., A.E. and H.S..

G.Y. and B.R.B contributed equally in this work.

The final version of the manuscript was approved by all authors.

## Reference

(1) Zhang, H.; Mielke, L.; Sychev, D.; Sun, N.; Hermes, I.; Schlicke, H.; Lissel, F. S.-C. Nanoparticles in Printed Photovoltaics: Materials, Methods and Potential. *Mater. Today Energy* **2025**, 102054. https://doi.org/10.1016/j.mtener.2025.102054.

(2) Ketelsen, B.; Tjarks, P. P.; Schlicke, H.; Liao, Y.-C.; Vossmeyer, T. Fully Printed Flexible Chemiresistors with Tunable Selectivity Based on Gold Nanoparticles. *Chemosensors* **2020**, *8* (4). https://doi.org/10.3390/chemosensors8040116.

(3) Cui, W.; Lu, W.; Zhang, Y.; Lin, G.; Wei, T.; Jiang, L. Gold Nanoparticle Ink Suitable for Electric-Conductive Pattern Fabrication Using in Ink-Jet Printing Technology. *Colloids Surf. A: Physicochem. Eng. Asp.* **2010**, *358* (1), 35–41. https://doi.org/10.1016/j.colsurfa.2010.01.023.

(4) Chen, S.; Pan, Q.; Wu, T.; Xie, H.; Xue, T.; Su, M.; Song, Y. Printing Nanoparticle-Based Isotropic/Anisotropic Networks for Directional Electrical Circuits. *Nanoscale* **2022**, *14* (40), 14956–14961. https://doi.org/10.1039/D2NR03892G.

(5) Garcia, M. A. Surface Plasmons in Metallic Nanoparticles: Fundamentals and Applications. *J. Phys. D: Appl. Phys.* **2011**, *44* (28), 283001. https://doi.org/10.1088/0022-3727/44/28/283001.

(6) Montaudo, G.; Puglisi, C.; Samperi, F. Primary Thermal Degradation Mechanisms of PET and PBT. *Polym. Degrad. Stab.* **1993**, *42* (1), 13–28. https://doi.org/10.1016/0141-3910(93)90021-A.

(7) Reiser, B.; González-García, L.; Kanelidis, I.; Maurer, J. H. M.; Kraus, T. Gold Nanorods with Conjugated Polymer Ligands: Sintering-Free Conductive Inks for Printed Electronics. *Chem. Sci.* **2016**, *7* (7), 4190–4196. https://doi.org/10.1039/C6SC00142D.

(8) Moon, H.; Lee, H.; Kwon, J.; Suh, Y. D.; Kim, D. K.; Ha, I.; Yeo, J.; Hong, S.; Ko, S. H. Ag/Au/Polypyrrole Core-Shell Nanowire Network for Transparent, Stretchable and Flexible Supercapacitor in Wearable Energy Devices. *Sci. Rep.* **2017**, *7* (1), 41981. https://doi.org/10.1038/srep41981.

(9) Jeon, J.-W.; Ledin, P. A.; Geldmeier, J. A.; Ponder, J. F. Jr.; Mahmoud, M. A.; El-Sayed, M.; Reynolds, J. R.; Tsukruk, V. V. Electrically Controlled Plasmonic Behavior of Gold Nanocube@Polyaniline Nanostructures: Transparent Plasmonic Aggregates. *Chem. Mater.* **2016**, *28* (8), 2868–2881. https://doi.org/10.1021/acs.chemmater.6b00882.

(10) Jeon, J.-W.; Zhou, J.; Geldmeier, J. A.; Ponder, J. F. Jr.; Mahmoud, M. A.; El-Sayed, M.; Reynolds, J. R.; Tsukruk, V. V. Dual-Responsive Reversible Plasmonic Behavior of Core–Shell Nanostructures with pH-Sensitive and Electroactive Polymer Shells. *Chem Mater.* **2016**, *28* (20), 7551–7563. https://doi.org/10.1021/acs.chemmater.6b04026.

(11) Jiang, N.; Shao, L.; Wang, J. (Gold Nanorod Core)/(Polyaniline Shell) Plasmonic Switches with Large Plasmon Shifts and Modulation Depths. *Adv. Mater.* **2014**, *26* (20), 3282–3289. https://doi.org/10.1002/adma.201305905.

(12) Lu, W.; Jiang, N.; Wang, J. Active Electrochemical Plasmonic Switching on Polyaniline-Coated Gold Nanocrystals. *Adv. Mater.* **2017**, *29* (8), 1604862. https://doi.org/10.1002/adma.201604862.

(13) Jones, A.; Searles, E. K.; Mayer, M.; Hoffmann, M.; Gross, N.; Oh, H.; Fery, A.; Link, S.; Landes, C. F. Active Control of Energy Transfer in Plasmonic Nanorod–Polyaniline Hybrids. *J. Phys. Chem. Lett.* **2023**, *14* (36), 8235–8243. https://doi.org/10.1021/acs.jpclett.3c01990.

(14) Yi, G.; Hoffmann, M.; Seçkin, S.; König, T. A. F.; Hermes, I.; Rossner, C.; Fery, A. Toward Coupling across Inorganic/Organic Hybrid Interfaces: Polyaniline-Coated Gold Nanoparticles with 4-Aminothiophenol as Gold-Anchoring Moieties. *Colloid Polym. Sci.* **2025**, *303* (9), 1743–1751. https://doi.org/10.1007/s00396-024-05262-x.

(15) Chiang, J.-C.; MacDiarmid, A. G. ‘Polyaniline’: Protonic Acid Doping of the Emeraldine Form to the Metallic Regime. *Synth. Met.* **1986**, *13* (1), 193–205. https://doi.org/10.1016/0379-6779(86)90070-6.

(16) Roncali, J.; Leriche, P.; Cravino, A. From One- to Three-Dimensional Organic Semiconductors: In Search of the Organic Silicon? *Adv. Mater.* **2007**, *19* (16), 2045–2060. https://doi.org/10.1002/adma.200700135.

(17) Xu, K.; Qin, L.; Heath, J. R. The Crossover from Two Dimensions to One Dimension in Granular Electronic Materials. *Nature Nanotech.* **2009**, *4* (6), 368–372. https://doi.org/10.1038/nnano.2009.81.

(18) Wang, Y.; Duan, C.; Peng, L.; Liao, J. Dimensionality-Dependent Charge Transport in Close-Packed Nanoparticle Arrays: From 2D to 3D. *Sci. Rep.* **2014**, *4* (1), 7565. https://doi.org/10.1038/srep07565.

(19) Zabet-Khosousi, A.; Dhirani, A.-A. Charge Transport in Nanoparticle Assemblies. *Chem. Rev.* **2008**, *108* (10), 4072–4124. https://doi.org/10.1021/cr0680134.

(20) Schmid, G.; Simon, U. Gold Nanoparticles: Assembly and Electrical Properties in 1–3 Dimensions. *Chem. Commun.* **2005**, 697–710. https://doi.org/10.1039/B411696H.

(21) Talapin, D. V; Lee, J.-S.; Kovalenko, M. V; Shevchenko, E. V. Prospects of Colloidal Nanocrystals for Electronic and Optoelectronic Applications. *Chem. Rev.* **2010**, *110* (1), 389–458. https://doi.org/10.1021/cr900137k.

(22) Kane, J.; Ong, J.; Saraf, R. F. Chemistry, Physics, and Engineering of Electrically Percolating Arrays of Nanoparticles: A Mini Review. *J. Mater. Chem.* **2011**, *21* (42), 16846–16858. https://doi.org/10.1039/C1JM12005K.

(23) Ghosh, S.; Bao, W.; Nika, D. L.; Subrina, S.; Pokatilov, E. P.; Lau, C. N.; Balandin, A. A. Dimensional Crossover of Thermal Transport in Few-Layer Graphene. *Nature Mater.* **2010**, *9* (7), 555–558. https://doi.org/10.1038/nmat2753.

(24) Faure, B.; Wetterskog, E.; Gunnarsson, K.; Josten, E.; Hermann, R. P.; Brückel, T.; Andreasen, J. W.; Meneau, F.; Meyer, M.; Lyubartsev, A.; Bergström, L.; Salazar-Alvarez, G.; Svedlindh, P. 2D to 3D Crossover of the Magnetic Properties in Ordered Arrays of Iron Oxide Nanocrystals. *Nanoscale* **2013**, *5* (3), 953–960. https://doi.org/10.1039/C2NR33013J.

(25) Qu, L.; Unruh, D.; Zimanyi, G. T. Percolative Charge Transport in Binary Nanocrystal Solids. *Phys. Rev. B* **2021**, *103* (19), 195303. https://doi.org/10.1103/PhysRevB.103.195303.

(26) Bascones, E.; Estévez, V.; Trinidad, J. A.; MacDonald, A. H. Electronic Correlations and Disorder in Transport through One-Dimensional Nanoparticle Arrays. *Phys. Rev. B* **2008**, *77* (24), 245422. https://doi.org/10.1103/PhysRevB.77.245422.

(27) Hanske, C.; Tebbe, M.; Kuttner, C.; Bieber, V.; Tsukruk, V. V; Chanana, M.; König, T. A. F.; Fery, A. Strongly Coupled Plasmonic Modes on Macroscopic Areas via Template-Assisted Colloidal Self-Assembly. *Nano Lett.* **2014**, *14* (12), 6863–6871. https://doi.org/10.1021/nl502776s.

(28) Schletz, D.; Schultz, J.; Potapov, P. L.; Steiner, A. M.; Krehl, J.; König, T. A. F.; Mayer, M.; Lubk, A.; Fery, A. Exploiting Combinatorics to Investigate Plasmonic Properties in Heterogeneous Ag-Au Nanosphere Chain Assemblies. *Adv. Optical Mater.* **2021**, *9* (9), 2001983. https://doi.org/10.1002/adom.202001983.

(29) Glatz, B. A.; Fery, A. The Influence of Plasma Treatment on the Elasticity of the in Situ Oxidized Gradient Layer in PDMS: Towards Crack-Free Wrinkling. *Soft Matter* **2019**, *15* (1), 65–72. https://doi.org/10.1039/C8SM01910J.

(30) Seçkin, S.; Singh, P.; Jaiswal, A.; König, T. A. F. Super-Radiant SERS Enhancement by Plasmonic Particle Gratings. *ACS Appl. Mater. Interfaces* **2023**, *15* (36), 43124–43134. https://doi.org/10.1021/acsami.3c07532.

(31) Yu, Z.; Sarkar, S.; Seçkin, S.; Sun, N.; Ghosh, A. K.; Wießner, S.; Zhou, Z.; Fery, A. 2D Wrinkle Assisted Zigzag Plasmonic Chains for Isotropic SERS Enhancement. *Sci. Rep.* **2025**, *15* (1), 3662. https://doi.org/10.1038/s41598-025-87504-8.

(32) Probst, P. T.; Mayer, M.; Gupta, V.; Steiner, A. M.; Zhou, Z.; Auernhammer, G. K.; König, T. A. F.; Fery, A. Mechano-Tunable Chiral Metasurfaces via Colloidal Assembly. *Nat. Mater.* **2021**, *20* (7), 1024–1028. https://doi.org/10.1038/s41563-021-00991-8.

(33) Steiner, A. M.; Mayer, M.; Seuss, M.; Nikolov, S.; Harris, K. D.; Alexeev, A.; Kuttner, C.; König, T. A. F.; Fery, A. Macroscopic Strain-Induced Transition from Quasi-Infinite Gold Nanoparticle Chains to Defined Plasmonic Oligomers. *ACS Nano* **2017**, *11* (9), 8871–8880. https://doi.org/10.1021/acsnano.7b03087.

(34) Lu, Y.; Lam, S. H.; Lu, W.; Shao, L.; Chow, T. H.; Wang, J. All-State Switching of the Mie Resonance of Conductive Polyaniline Nanospheres. *Nano Lett.* **2022**, *22* (3), 1406–1414. https://doi.org/10.1021/acs.nanolett.1c04969.

(35) Nečas, D.; Klapetek, P. Gwyddion: An Open-Source Software for SPM Data Analysis. *Cent. Eur. J. Phys.* **2012**, *10*(1), 181–188. https://doi.org/10.2478/s11534-011-0096-2.

(36) Chen, H.; Kou, X.; Yang, Z.; Ni, W.; Wang, J. Shape- and Size-Dependent Refractive Index Sensitivity of Gold Nanoparticles. *Langmuir* **2008**, *24* (10), 5233–5237. https://doi.org/10.1021/la800305j.

(37) Bartolucci, S. F.; Leff, A. C.; Maurer, J. A. Gold–Copper Oxide Core–Shell Plasmonic Nanoparticles: The Effect of pH on Shell Stability and Mechanistic Insights into Shell Formation. *Nanoscale Adv.* **2024**, *6* (9), 2499–2507. https://doi.org/10.1039/D3NA01000G.

(38) Yu, Y.; Ng, C.; König, T. A. F.; Fery, A. Tackling the Scalability Challenge in Plasmonics by Wrinkle-Assisted Colloidal Self-Assembly. *Langmuir* **2019**, *35* (26), 8629–8645. https://doi.org/10.1021/acs.langmuir.8b04279.

(39) Bhadra, S.; Khastgir, D. Degradation and Stability of Polyaniline on Exposure to Electron Beam Irradiation (Structure–Property Relationship). *Polym. Degrad. Stab.* **2007**, *92* (10), 1824–1832. https://doi.org/10.1016/j.polymdegradstab.2007.07.004.

(40) Afzal, A. B.; Akhtar, M. J.; Nadeem, M.; Hassan, M. M. Investigation of Structural and Electrical Properties of Polyaniline/Gold Nanocomposites. *J. Phys. Chem. C* **2009**, *113* (40), 17560–17565. https://doi.org/10.1021/jp902725d.

(41) Vasileva, A. A.; Mamonova, D. V; Mikhailovskii, V.; Petrov, Y. V; Toropova, Y. G.; Kolesnikov, I. E.; Leuchs, G.; Manshina, A. A. 3D Nanocomposite with High Aspect Ratio Based on Polyaniline Decorated with Silver NPs: Synthesis and Application as Electrochemical Glucose Sensor. *Nanomaterials* **2023**, *13* (6). https://doi.org/10.3390/nano13061002.

(42) Tseng, R. J.; Baker, C. O.; Shedd, B.; Huang, J.; Kaner, R. B.; Ouyang, J.; Yang, Y. Charge Transfer Effect in the Polyaniline-Gold Nanoparticle Memory System. *Appl. Phys. Lett.* **2007**, *90* (5), 053101. https://doi.org/10.1063/1.2434167.

(43) Tseng, R. J.; Huang, J.; Ouyang, J.; Kaner, R. B.; Yang. Polyaniline Nanofiber/Gold Nanoparticle Nonvolatile Memory. *Nano Lett.* **2005**, *5* (6), 1077–1080. https://doi.org/10.1021/nl050587l.

(44) Zhang, X.; Goux, W. J.; Manohar, S. K. Synthesis of Polyaniline Nanofibers by “Nanofiber Seeding.” *J. Am. Chem. Soc.* **2004**, *126* (14), 4502–4503. https://doi.org/10.1021/ja031867a.

(45) Mallick, K.; Witcomb, M. J.; Scurrell, M. S. Gold in Polyaniline: Recent Trends. *Gold Bull.* **2006**, *39* (4), 166–174. https://doi.org/10.1007/BF03215550.

(46) Mushibe, E. K.; Murphy, S. C.; Raiti-Palazzolo, K.; Mccarthy, D. L.; Obuya, E. A.; Chiguma, J.; Jones, W. E. Dielectric Nanowire Composites: One-Pot Synthesis of Gold Nanoparticles Encapsulated in Polyaniline Fibers. *MRS Online Proceedings Library* **2012**, *1453* (1), 93–98. https://doi.org/10.1557/opl.2013.1133.

(47) Babu, V. J.; Vempati, S.; Ramakrishna, S. Conducting Polyaniline-Electrical Charge Transportation. *Materials Sciences and Applications* **2013**, *04* (01). https://doi.org/10.4236/msa.2013.41001.

(48) Gupta, K.; Jana, P. C.; Meikap, A. K. Electrical Transport and Optical Properties of the Composite of Polyaniline Nanorod with Gold. *Solid State Sci.* **2012**, *14* (3), 324–329. https://doi.org/10.1016/j.solidstatesciences.2011.12.003.

(49) Kim, K. H.; Lara-Avila, S.; He, H.; Kang, H.; Hong, S. J.; Park, M.; Eklöf, J.; Moth-Poulsen, K.; Matsushita, S.; Akagi, K.; Kubatkin, S.; Park, Y. W. Probing Variable Range Hopping Lengths by Magneto Conductance in Carbonized Polymer Nanofibers. *Sci. Rep.* **2018**, *8* (1), 4948. https://doi.org/10.1038/s41598-018-23254-0.

# Nanoscale Charge Transport in Au@PANI Assemblies: Bulk-like Films and Linear Assemblies

*Gyusang Yi [a,§], Borja Rodriguez-Barea [c,d,§], Gabriele Carelli [a], Lukas Mielke [a], Andreas Fery [a,b], Artur Erbe [c,d,*], Hendrik Schlicke [a,*]*

[a] Leibniz Institute of Polymer Research Dresden, 01069 Dresden, Germany.

[b] Chair of Physical Chemistry of Polymeric Materials, Dresden University of Technology, 01062 Dresden, Germany

[c] Institute of Ion Beam Physics and Materials Research, Helmholtz-Zentrum, Dresden-Rossendorf, 01328 Dresden, Germany

[d] Institute of semiconductors and microsystems, Dresden University of Technology, 01062 Dresden, Germany

§ contributed equally to this work

* Corresponding author
E-mail: a.erbe@hzdr.de, schlicke@ipfdd.de

## Supporting information

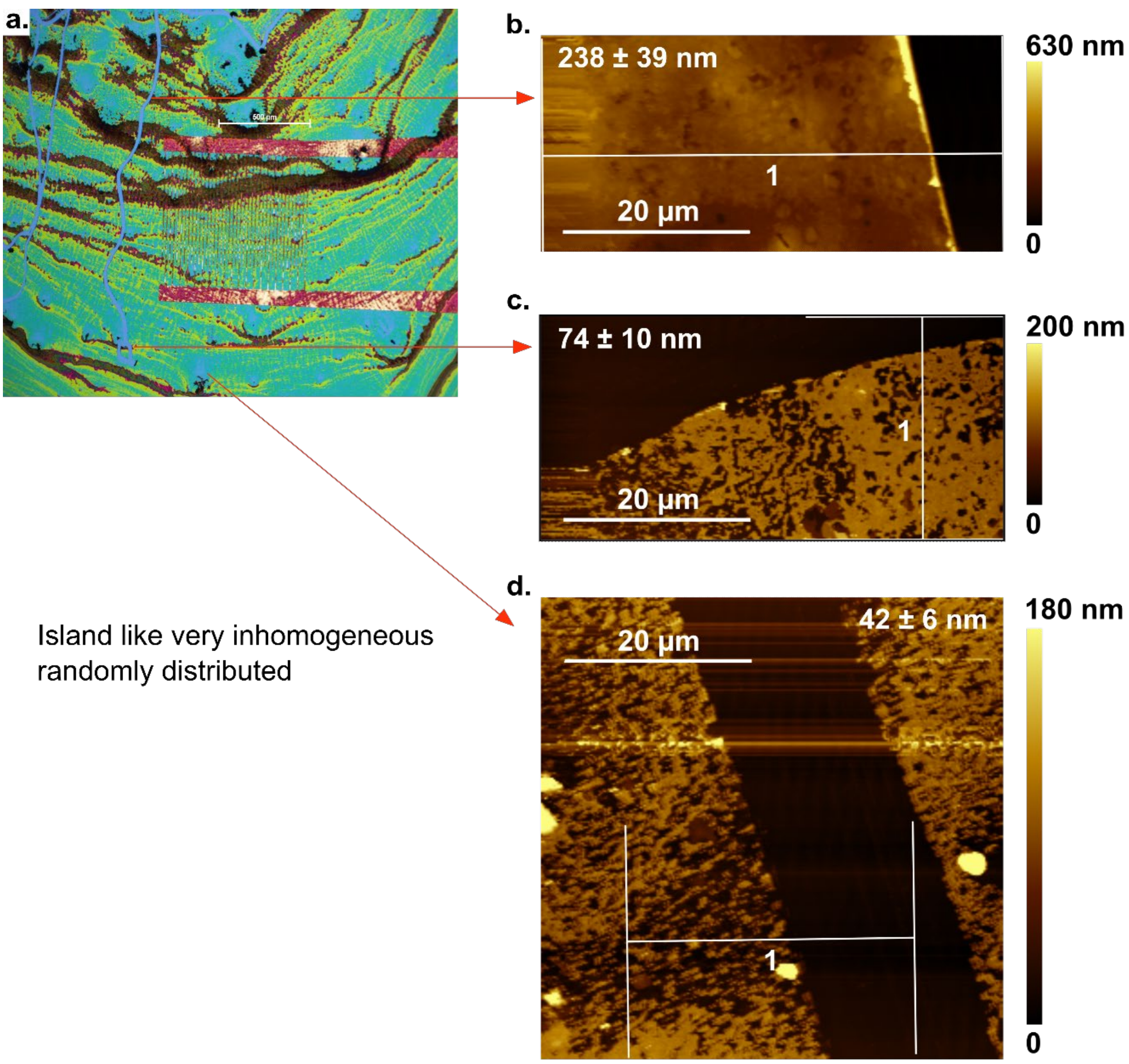

**Fig. S1.** (a) Optical Micrographs of representative bulk-like (inhomogeneous and discontinuous) films by optical microscope and AFM. (b-d) The arrows show the approximate position of the AFM scans.

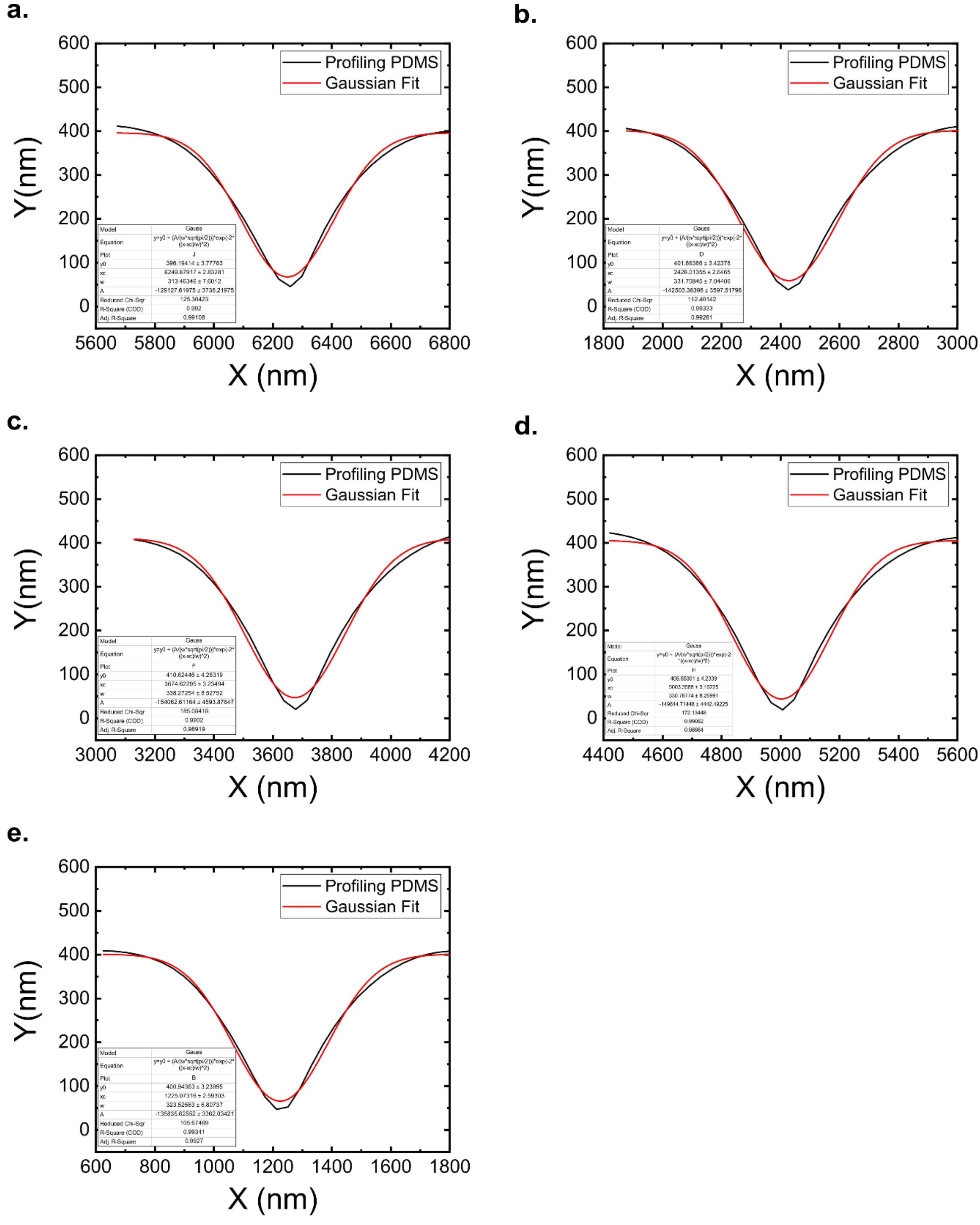

**Fig. S2.** Profiling of PDMS wrinkles and Gaussian fitting of FWHM and Height of PDMS wrinkles shown in Fig. 3 b)

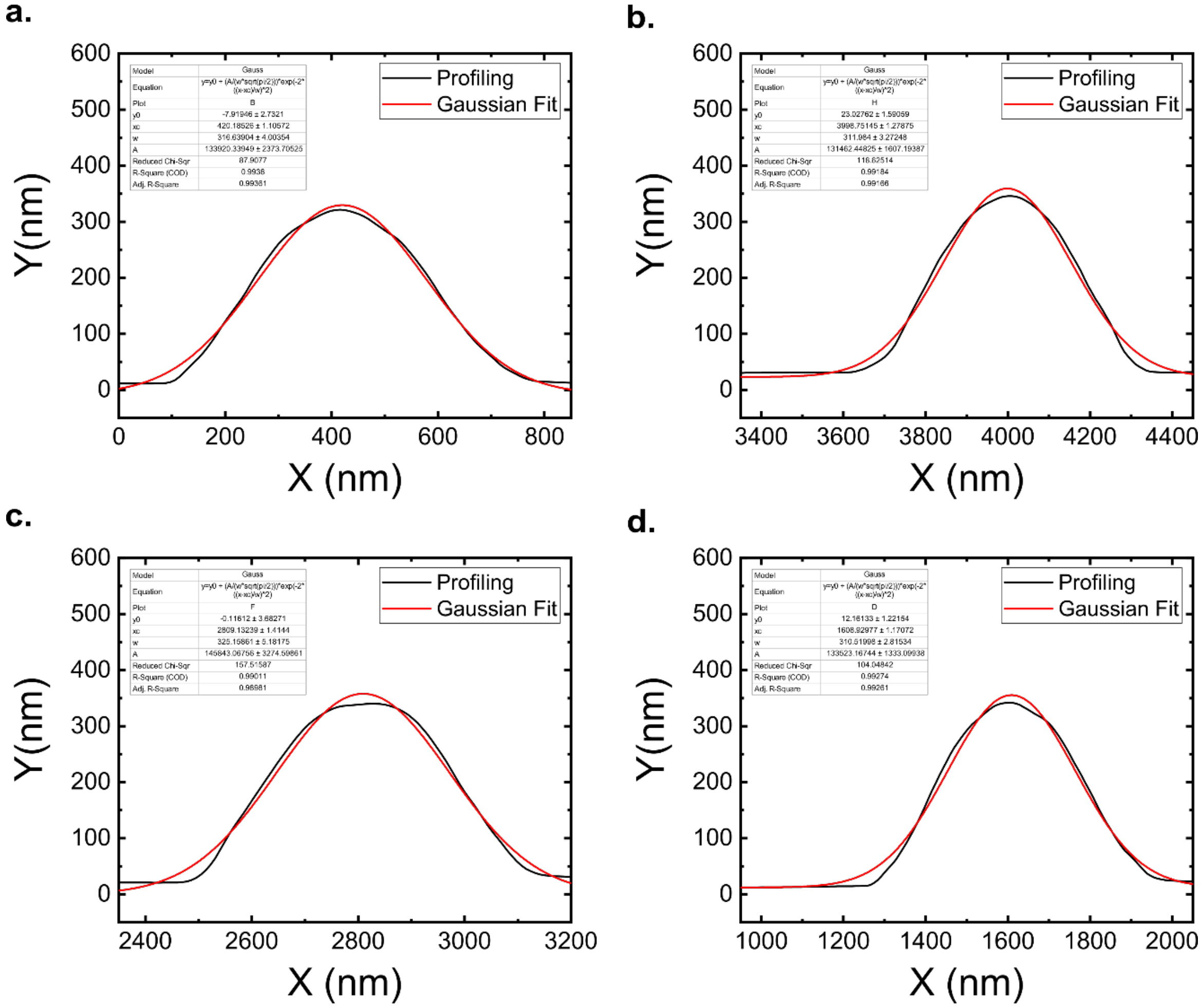

**Fig. S3.** Particle line profiling and Gaussian fitting for FWHM and height (line assemblies shown in Fig. 3d.)

.

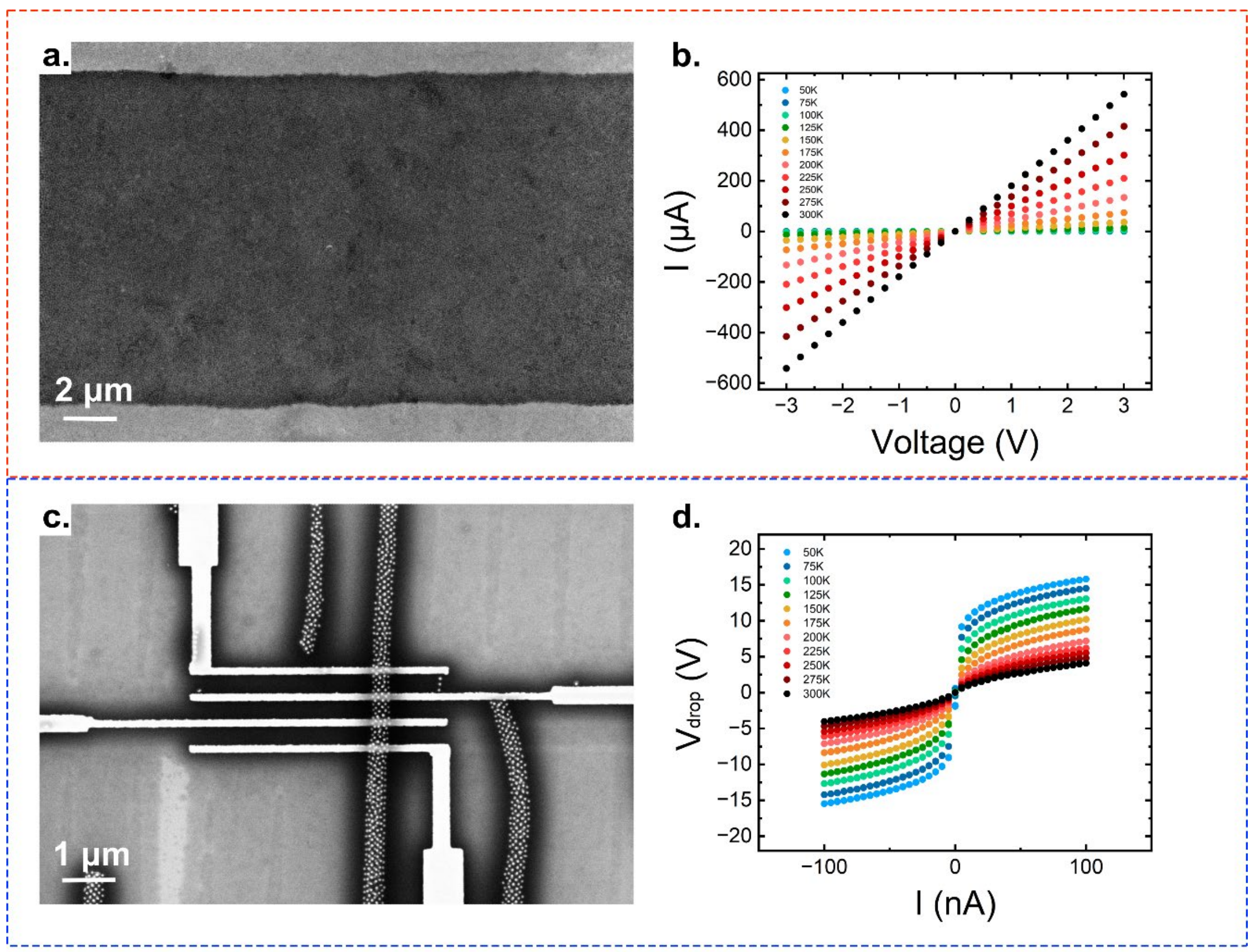

**Fig. S4.** A second electrode measurement to confirm the reproducibility of the systems discussed in the main text. a) An SEM image of the Au@PANI 3D bulk-like film in a section of the IDE electrodes, b) the two-probe electrical characterization across the 50–300 K temperature range (red box) c) an SEM image of the PDMS-template-assisted Au@PANI line assemblies positioned on top of four bottom electrodes, and d) their four-probe electrical characterization in the 50–300 K temperature range.

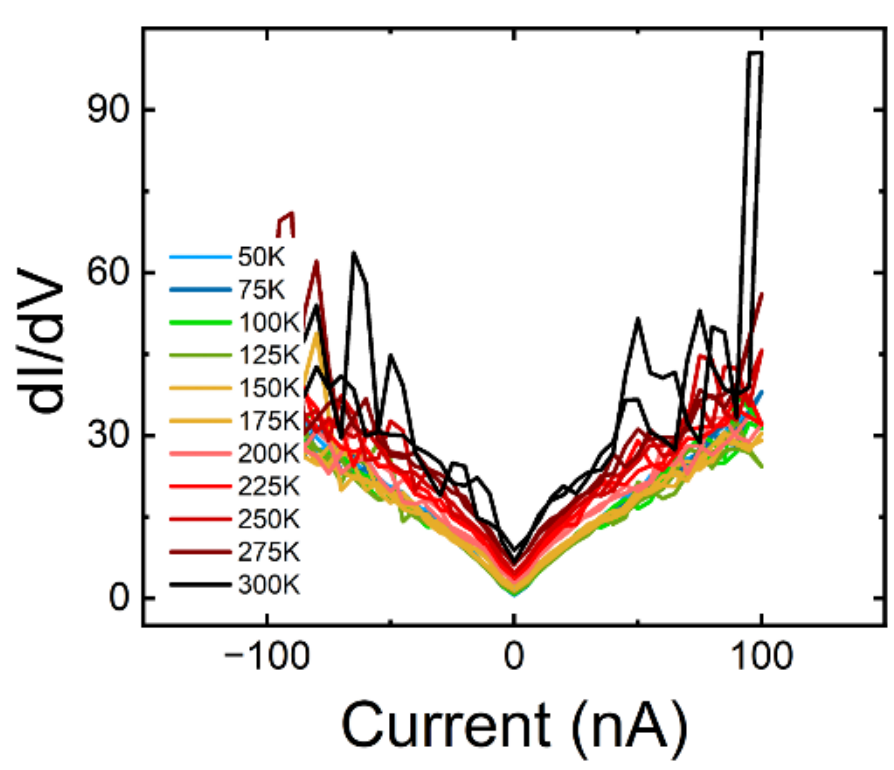

**Fig. S5.** Differential conductance of the linear chain assembly shown in Fig. S4c from a four-terminal configuration at different temperatures ranging from 50K–300K.

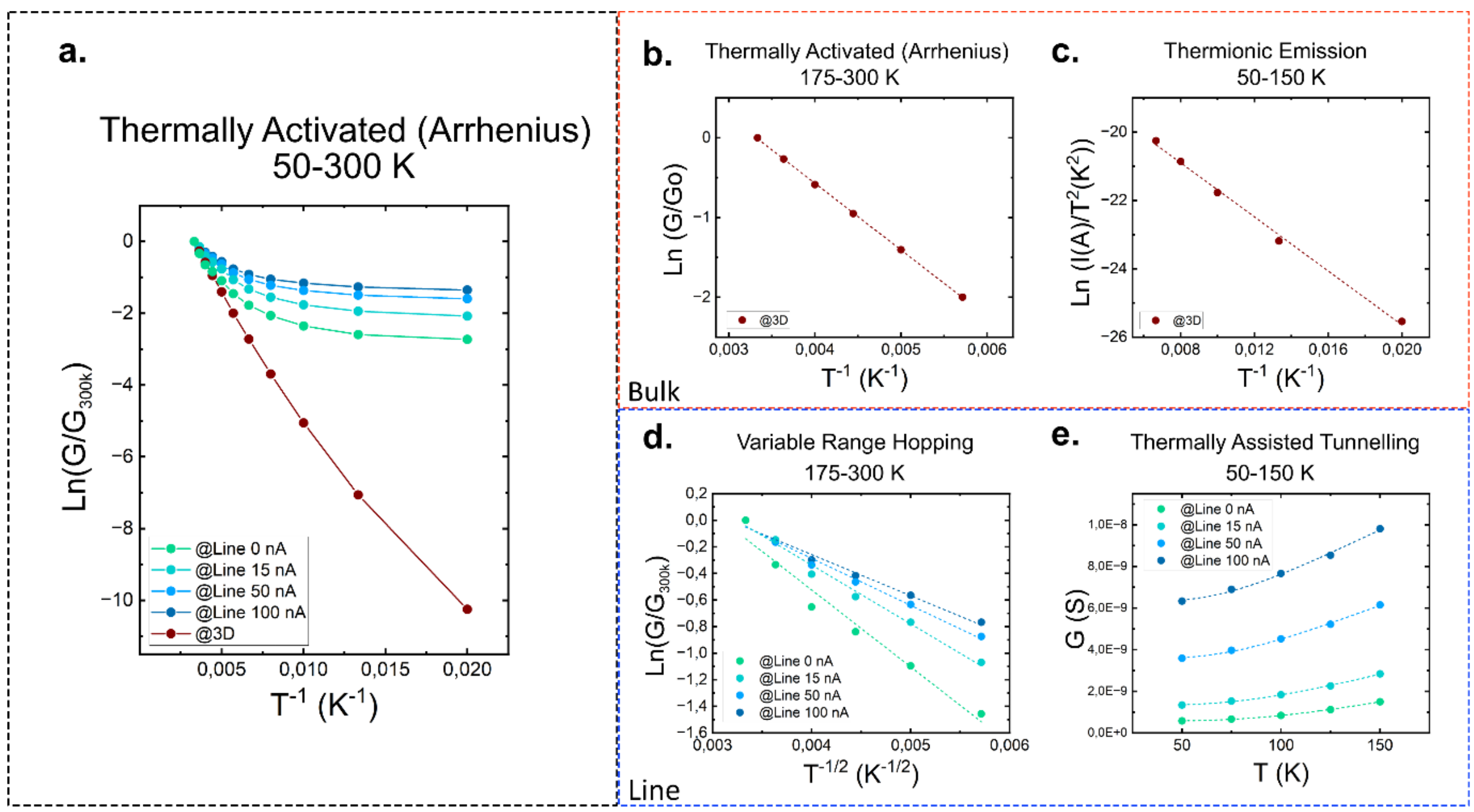

**Fig. S6.** Fittings for a second electrode systems presented in Fig. S4: a) Logarithm of the normalized conductance plotted against $T^{-1}$ for linearization according to the Arrhenius mechanism across the 50–300 K temperature range for both line and bulk-like systems. The solid line serves as a visual guide. The top row (red box) shows the fittings for the bulk-like film: b) Arrhenius plot restricted to the high-temperature regime (175–300 K), with a linear fit shown as a solid dashed line. c) Displays thermionic conduction mechanism in the low-temperature regime (50–150 K), with a linear fit shown as a solid dashed line. The bottom row (blue box) shows the fittings for the line assembly: d) 1D VRH model in the high-temperature regime (175–300 K), with a linear fit shown as a solid dashed line. e) Thermally Assisted Tunneling (TAT) model in the low-temperature regime (50–150 K), with a linear fit shown as a solid dashed line.

**Table S1.** Parameters obtained from fitting the Thermally Assisted Tunneling (TAT) model for both the system discussed in the main text and the replicated system presented in the Supporting Information.

| | **Input current** | $\mathbf{G_{base}}$ | $\mathbf{G_1}$ | $\mathbf{E_a}$ **(mV)** |
|---|---|---|---|---|
| **Main Text** | 100 nA | 5.8 nS | 16 nS | 24.2 |
| | 50 nA | 3.3 nS | 13 nS | 25.7 |
| | 15 nA | 1.3 nS | 11 nS | 30.0 |
| | 0 nA | 5.4 nS | 11 nS | 36.1 |
| **APPENDIX** | 100 nA | 6.3 nS | 23 nS | 24.8 |
| | 50 nA | 3.6 nS | 19 nS | 26.0 |
| | 15 nA | 1.3 nS | 13 nS | 28.7 |
| | 0 nA | 5.8 nS | 11 nS | 32.6 |

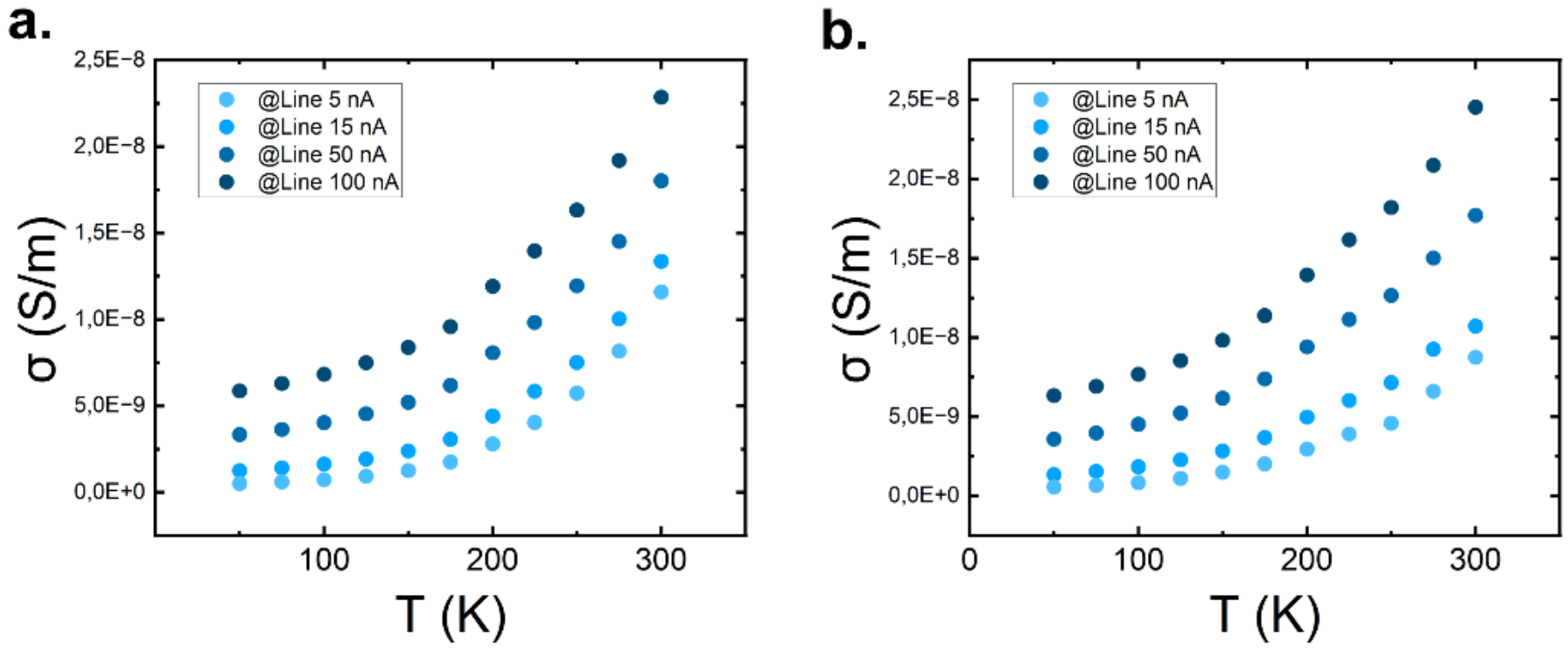

**Fig. S7.** Calculated conductivity for the linear assemblies. a) Assembly discussed in the main text. b) A second electrode presented in the Supporting Information.

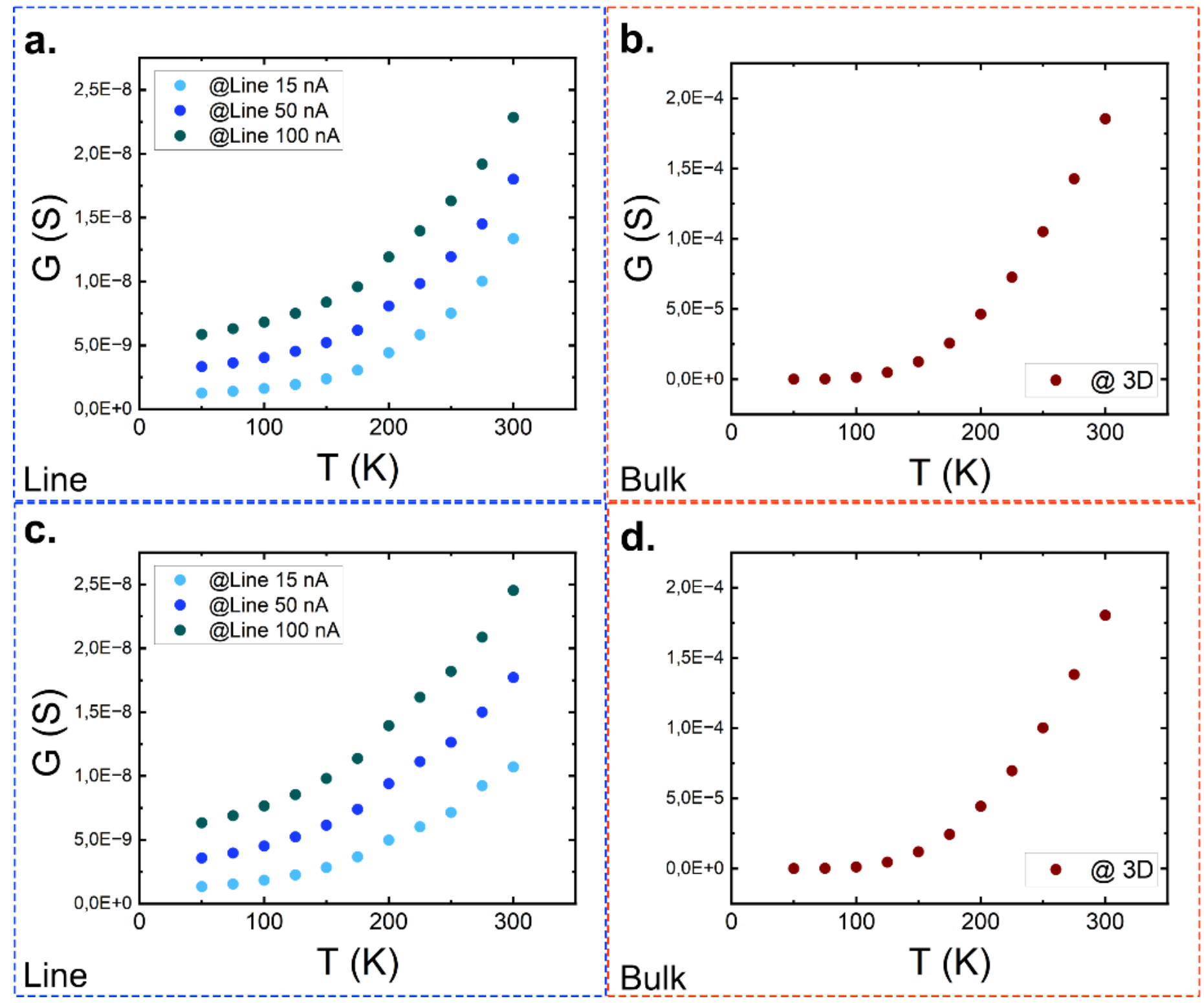

**Fig. S8.** Measured conductance versus temperature for all four systems. The top row shows the systems discussed in the main text, while the bottom row shows the other assembly systems presented in the Supporting Information. (a–d) Temperature-dependent conductance over the 50–300 K range.

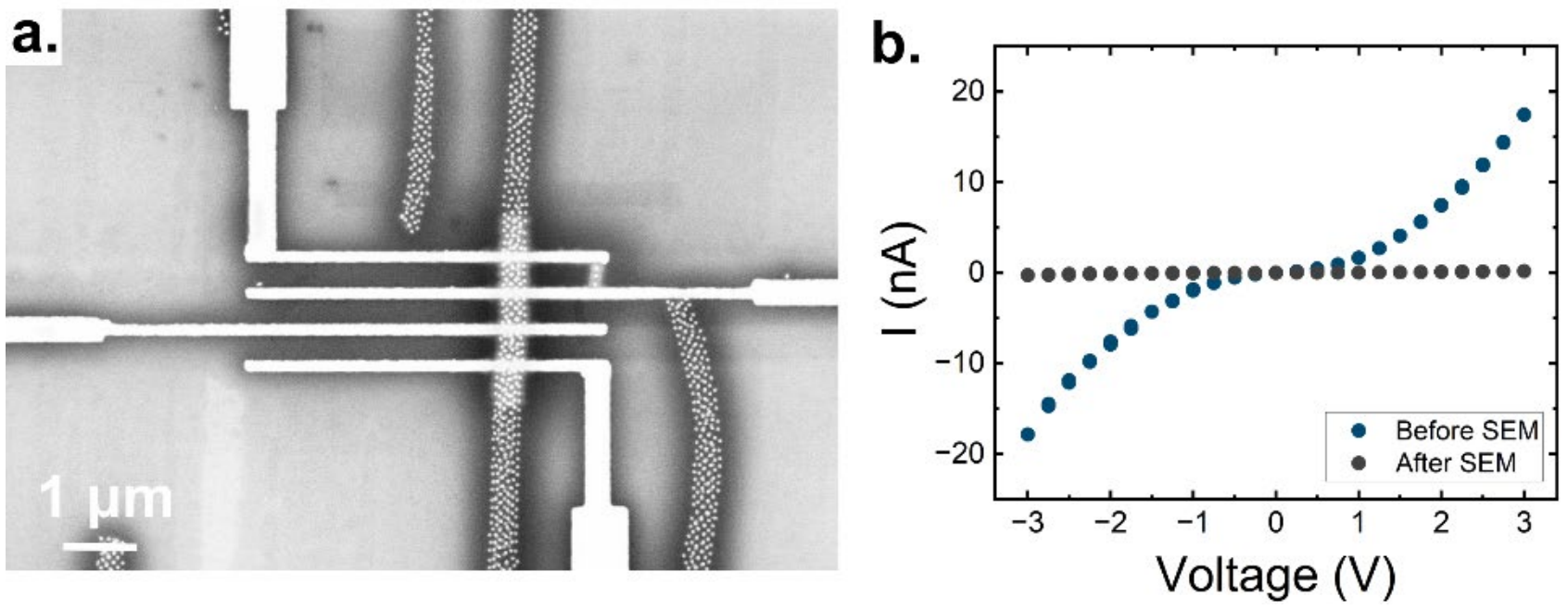

**Fig. S9**. a) SEM image of the damaged PANI after high resolution E-beam exposure (10 kV). b) Electrical characterization for the internal contacts before and after the E-beam exposure.

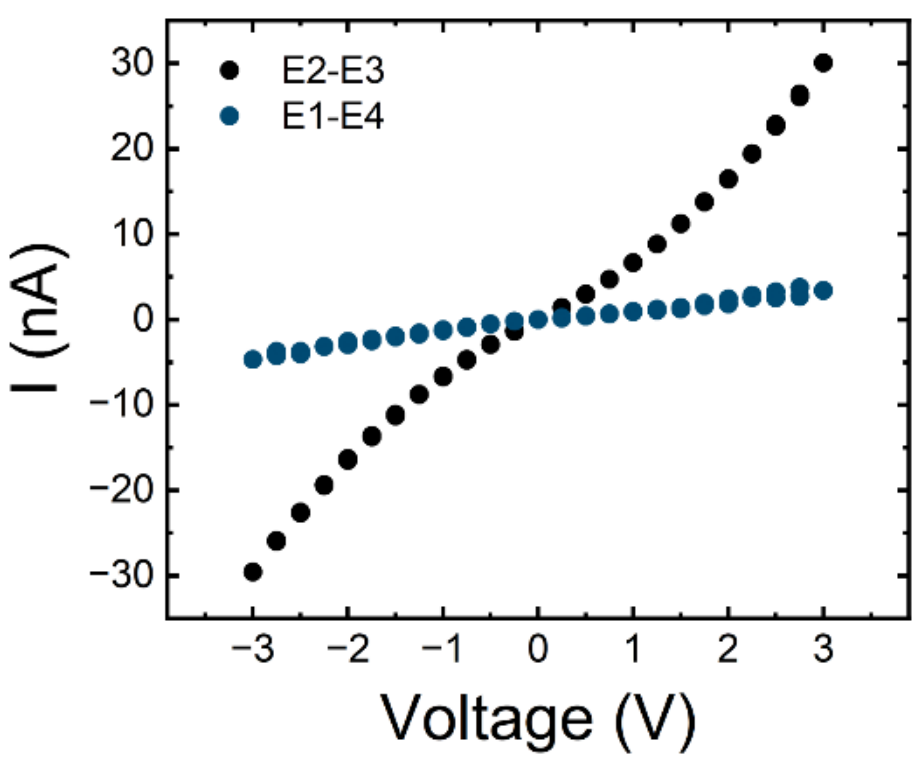

**Fig. S10.** Two-terminal measurements at room temperature confirming stable electrical connections for both the internal electrodes (E2–E3) and the external electrodes (E1–E4).